\documentclass{aa}  

\usepackage{graphicx}
\usepackage{txfonts}

\usepackage[colorlinks=true,citecolor=blue,linkcolor=blue]{hyperref}

\begin{document} 




\title{
Inferring redshift and galaxy properties via a multi-task neural net with probabilistic outputs}

\subtitle{An application to simulated MOONS spectra}

   \author{Michele Ginolfi \inst{1,2}
          \and
           Filippo Mannucci \inst{2}
          \and
          Francesco Belfiore \inst{2}
          \and
          Alessandro Marconi \inst{1,2}
          \and
          Nicholas Boardman \inst{3}
           \and 
           \newline
          Lucia Pozzetti \inst{4}
          \and
          Micol Bolzonella \inst{4}
         \and
          Enrico Di Teodoro \inst{1,2}
          \and
          Giovanni Cresci \inst{2}
          \and
          Vivienne Wild \inst{3}
          \and
          Myriam Rodrigues \inst{5}
          \and
          Roberto Maiolino \inst{6,7,8}
          \and
          Michele Cirasuolo \inst{9}
          \and
          Ernesto Oliva \inst{2}
          }

    \institute{
        Dipartimento di Fisica e Astronomia, Università di Firenze, Via G. Sansone 1, I-50019, Sesto F.no (Firenze), Italy
        \and
        INAF — Osservatorio Astrofisico di Arcetri, Largo E. Fermi 5, I-50125, Florence, Italy
        \and
        School of Physics and Astronomy, University of St Andrews, North Haugh, St Andrews KY16 9SS, UK
        \and
        INAF-Osservatorio di Astrofisica e Scienza dello Spazio di Bologna, Via Piero Gobetti 93/3, 40129 Bologna, Italy
        \and
        GEPI, Observatoire de Paris, PSL University, CNRS, France
        \and
        Cavendish Laboratory, University of Cambridge, 19 J. J. Thomson Ave., Cambridge CB3 0HE, UK
        \and
        Kavli Institute for Cosmology, University of Cambridge, Madingley Road, Cambridge CB3 0HA, UK
        \and
        Department of Physics and Astronomy, University College London, Gower Street, London WC1E 6BT, UK
        \and
        European Southern Observatory, Karl-Schwarzschild-Strasse 2, D-85748 Garching bei Muenchen, Germany
    }

\date{~}

\abstract{
The era of large-scale astronomical surveys demands innovative approaches for rapid and accurate analysis of extensive spectral data, and a promising direction in which to address this challenge is offered by machine learning.
Here, we introduce a new pipeline, \texttt{M-TOPnet} (Multi-Task network Outputting Probabilities), which employs a convolutional neural network with residual learning to simultaneously derive redshift and other key physical properties of galaxies from their spectra.  
Our tool efficiently encodes spectral information into a latent space, employing distinct downstream branches for each physical quantity, thereby benefiting from multi-task learning. 
Notably, our method handles the redshift output as a probability distribution, allowing for a more refined and robust estimation of this critical parameter. 
We demonstrate preliminary results using simulated data from the MOONS instrument, which will soon be operating at the ESO/VLT. 
We highlight the effectiveness of our tool in accurately predicting the redshift, stellar mass, and star formation rate of galaxies at $z \gtrsim 1-3$, even for faint sources ($m_H \sim 24$) 
for which traditional methods often struggle. 
Through analysis of the output probability distributions, we demonstrate that our pipeline enables robust quality screening of the results, achieving accuracy rates of up to 99\% in redshift determination (defined as predictions within $|\Delta z| < 0.01$ relative to the true redshift) with $8~h$ exposure spectra, while automatically identifying potentially problematic cases.
Our pipeline thus emerges as a powerful solution for the upcoming challenges in observational astronomy, combining precision, interpretability, and efficiency, all aspects that are crucial for analysing the massive datasets expected from next-generation instruments.
}

\keywords{Methods: machine learning, data analysis -- 
   Galaxies: evolution, ISM, high-redshift}

   \maketitle

\section{Introduction}
\label{sec:intro}

Spectra encode most of the information on the physical properties of galaxies; therefore, spectroscopy is an essential technique to probe the physical mechanisms that drive the formation and evolution of galaxies across cosmic time.
The analysis of emission lines, absorption lines, and continuum in spectra across ultraviolet (UV), optical, and infrared (IR) wavelengths allows one to infer information about the internal structure and various physical properties of galaxies (see comprehensive reviews by \citealp{Conroy2013}, \citealp{Maiolino2019} and \citealp{Kewley2019}).
By applying stellar population synthesis models to the entire galaxy spectrum, it is possible to constrain the recent star formation history of galaxies (\citealp{Madau1998, Bruzual2003, Cid2005,Maraston2005, Tojeiro2007, Carnall2018}). 
Additionally, spectral energy distributions (SEDs) provide insights into key quantities like stellar masses ($M_{\rm star}$), star formation rates (SFRs), and dust extinction (\citealp{Brinchmann2004, Conroy2013, Boquien2019}). 
Likewise, the analysis of key emission lines in the UV and optical spectra are particularly revealing: indeed, hydrogen lines, such as Lyman-alpha (Ly$\alpha$) in the UV and recombination lines like H$\alpha$ and H$\beta$ in the optical range, effectively trace the ionising photon rate emitted by young, massive stars (\citealp{Kennicutt1998b, Buat2002, Kennicutt2012, Matthee2023}). On the other hand, bright oxygen lines, including [OII] and [OIII], serve as powerful diagnostics for understanding the physical conditions within the ionised interstellar medium (\citealp{Stasinska2003,Jones2015, Berg2016, Sanders2023}).
Moreover, the measurement of certain optical line ratios — for example, involving [OII], [OIII], [NII], and H$\alpha$ — enables the determination of gas-phase metallicity via so-called strong-line methods (\citealp{Nagao2006, Mannucci2009, Steidel2014, Maiolino2019, Curti2023, Nakajima2023}). Stellar metallicity, on the other hand, is typically assessed through absorption lines found in stellar spectra, such as those of iron and magnesium (\citealp{Tantalo2004, Gallazzi2005}). These two types of  measurements offer complementary perspectives on the chemical evolution of galaxies (\citealp{Peng2015,Lian2018, Fraser2022, Looser2024}). 

A crucial piece of information gained from galaxy spectra is their redshift.
Accurate spectroscopic redshifts not only anchor galaxies in both space and time but also allow for precise calculations of rest-frame properties like colour and luminosity (e.g. \citealp{Alam2017, Schlafly2023}).
More importantly, redshift is key to tracking the evolutionary trends over cosmic time of physical properties like the above-mentioned luminosity, $M_{\rm star}$, metallicity, and the SFR (a notable example is the time evolution of the SFR density; see e.g. \citealp{Madau2014}), as well as the time evolution of scaling relations like the galaxy main sequence (e.g. \citealp{Speagle2014, Renzini2015, Rinaldi2024}), the mass-metallicity relation (e.g. \citealp{Tremonti2004, Graziani2020, Langan2023, Heintz2023}), and the fundamental metallicity relation (e.g. \citealp{Mannucci2010}).
This information can serve as a benchmark for calibrating both analytical and numerical models of galaxy formation and evolution (see the reviews by \citealp{Somerville2015} and \citealp{Crain2023}), allowing us to refine our understanding of the complex physical processes involved in the galaxy baryon cycle, including galaxy feedback and mechanisms of galaxy quenching (see e.g. \citealp{Piotrowska2022, Kurinchi-Vendhan2023}), the interplay between gas inflows and outflows driven by active galactic nuclei (AGNs) and supernovae (see e.g. \citealp{Dave2011, AnglesAlcazar2017}), and the intricate balance between star formation efficiency and metal and dust enrichment (see e.g. \citealp{Ginolfi2018, Ginolfi2020, Graziani2020, DiCesare2023}). 
\\\\
However, despite the vast amounts of spectral data produced by large surveys, extracting physical quantities from these complex datasets in an accurate and quick way remains a challenge (e.g. \citealp{Huertas-Company2023, Zhong2024, Iglesias-Navarro2024}).  
Traditional methods, such as those implemented in tools like \textit{redmonster} (\citealp{Hutchinson2016}) and \textit{redrock} (\citealp{Lan2023}), utilise cross-correlation (\citealp{Tonry1979}) and template-fitting techniques to process spectral data and derive redshift and physical properties (see e.g. \citealp{Bautista2018, Napolitano2023}).
Although these traditional methods are reliable, they tend to be computationally intensive and often struggle with low signal-to-noise ratios and sky subtraction residuals, leading to increased failure rates and inaccuracies with low-quality data (\citealp{Bolton2012, Zhong2024}). 
Moreover, the efficiency of these techniques can be severely limited by the need for extensive template libraries or optimal initial guesses to ensure accuracy, making them less scalable for the vast data volumes expected from forthcoming astronomical surveys  (see e.g. \citealp{Newman2013}). Therefore, traditional methods will soon become a prohibitive bottleneck.

Over the next half-decade, large-multiplicity instruments like the Dark Energy Spectroscopic Instrument (DESI; \citealp{DESI2022, Hahn2023}), 
the 4-metre Multi-Object Spectroscopic Telescope (4MOST; \citealp{deJong2014, deJong2019}), 
the Subaru Prime Focus Spectrograph (PFS; \citealp{Tamura2022}),
and the Multi-Object Optical and Near-infrared Spectrograph (MOONS; \citealp{Cirasuolo2020, Maiolino2020}) will capture spectra from billions of galaxies, which would require tens or hundreds of billions of CPU hours for analysis (see e.g. \citealp{Huertas-Company2023}).
To optimise the scientific outcomes of these huge datasets, strategies to perform fast, efficient, and accurate automated analyses become mandatory.
Constructing fully data-driven models and employing 
deep learning based methods offers advantages in terms of efficiency, scalability, and flexibility, making them ideal for processing the enormous volumes of data anticipated from future spectroscopic surveys.
\\\\
For these reasons, it is not surprising that machine learning (ML) -based models are becoming increasingly prominent in the field of astronomy  using both unsupervised and supervised learning methods, among which artificial neural networks (ANNs) are very popular (see e.g. three recent reviews by \citealp{Baron2019}, \citealp{Smith2023}, and  \citealp{Huertas-Company2023} for a comprehensive overview of the state-of-the art of ML in astronomy and insights into the functioning of these popular methods).
In the field of galaxy formation and evolution, these techniques have been particularly successful in vision tasks involving galaxy images.
For instance, convolutional neural networks (CNN) have been successfully employed for galaxy morphology classification (e.g. \citealp{Dieleman2015, Huertas-Company2015, Ghosh2020, Walmsley2022}),  the identification of merging systems (e.g. \citealp{Ciprijanovic2020, Bickley2021, Ferreira2024}), and the detection of strong gravitational lenses (e.g. \citealp{Lanusse2018, Jacobs2019}).
Beyond classification, ML, and especially ANNs, has been applied to galaxy images to infer physical properties like photometric redshifts (e.g. \citealp{Collister2004, Pasquet2019, Zhou2021}) and resolved stellar populations (e.g. \citealp{Buck2021}), and to perform segmentation tasks like deblending (e.g. \citealp{Boucaud2020, Melchior2021}) and pixel-level morphological classifications using encoder-decoder architectures (e.g. \citealp{Hausen2020}).

From a spectroscopy perspective, while the integration of ML in spectral data analysis is also gaining traction, it is still less common than image-based applications, likely due to the higher complexity of spectral data and the established reliability of classical methods.
Some works have focussed on classifying galaxy spectra and estimating spectroscopic redshifts using deep CNNs (e.g. \citealp{Stivaktakis2018}) or Bayesian networks (e.g. \citealp{Podsztavek2022}) to account for uncertainties, reaching human-expert precision in localising and classifying spectral features (e.g. \citealp{Busca2018}), and outperforming standard methods in terms of speed and accuracy for redshift prediction (e.g. \citealp{Zhong2024}).
Moreover, ANNs and other supervised ML techniques (like AdaBoost with Decision Trees; see e.g. \citealp{Ucci2017}) have been used to determine key physical properties of galaxies from their emission line spectra, such as density, metallicity, and the ionisation parameter in the ISM, training the models with a large library of synthetic spectra (e.g. \citealp{Ucci2018}) or with labels obtained through common full spectrum fitting routines (e.g. \citealp{Wang2024}).
Also, probabilistic ML and simulation-based inference have been used to estimate the star formation histories of galaxies from their optical absorption spectra (see \citealp{Iglesias-Navarro2024}), and unsupervised learning techniques like auto-encoders have proven successful in reliably capturing spectral features, providing highly realistic reconstructions for galaxy spectra from an interpretable latent space, and detecting outliers (see \citealp{Melchior2023, Liang2023}).
\\

Capitalising on these advancements in spectroscopic analysis and borrowing techniques from the broader ML research community, we have developed a novel pipeline, which we dub \texttt{M-TOPnet} (Multi-Task network Outputting Probabilities) to facilitate reading and for simplicity.
\texttt{M-TOPnet} employs a CNN with residual learning (\citealp{He2015}; see applications to astronomical spectra in \citealp{Li2018}, \citealp{Zhong2024}, and \citealp{Moradi2024}) designed to simultaneously derive redshift, emission or absorption line locations, and key physical properties of galaxies (currently tested for $M_{\rm star}$ and SFR) from their spectra within a unified framework.
Our tool efficiently encodes spectral information (both from lines and continuum) into a shared embedding space. Distinct branches for each physical quantity depart from this common representation, leveraging the benefits of multi-task learning (e.g. \citealp{Caruana1997, Ruder2017, Crawshaw2020}).

To address the critical aspect of prediction uncertainty, we followed a two-fold approach. First, we formulated the redshift prediction as a classification task by mapping real redshift values into small bins determined by the spectral resolution of our spectra (a similar approach has been tested by e.g. \citealp{Carrasco2013, Stivaktakis2018, Stewart2022, Pankaj2022}). This method accounts for aleatoric uncertainty (uncertainty due to inherent noise in the data), allowing for the possibility of training the model with a probability distribution function (PDF) of the redshift (whose measurement can be influenced by the quality of the spectrum), and thus obtaining a PDF for the prediction.
Second, we experimented with incorporating Monte Carlo (MC) dropout to account for epistemic uncertainty (uncertainty due to model parameters) and reconstruct the posterior distribution function by averaging multiple inferences.
Monte Carlo dropout involves randomly  dropping units during each forward pass at inference time (see a discussion in Section \ref{sec:methods}), simulating a form of Bayesian approximation to estimate model uncertainty
(\citealp{Gal2015}; see also astronomical applications in \citealp{Ferreira2020,Perreault2017, Leung2019, Podsztavek2022}).

A key feature of our methodology is its interpretability. By analysing the embedding layers, we demonstrate the ability of \texttt{M-TOPnet} to learn and distinguish between important, yet unlabelled, galaxy characteristics, such as differentiating between star-forming and passive galaxies.

We present preliminary results using simulated data for the fibre-fed multi-object spectrograph MOONS (\citealp{Cirasuolo2020}), which will soon begin operations at the Very Large Telescope (VLT).
MOONS represents a transformative leap in our ability to study galaxy evolution, particularly around the epoch of cosmic noon ($z  \sim 1-2.5$). Its unique combination of large multiplexing (1000 fibers), high sensitivity, broad simultaneous spectral coverage extending in the near-IR ($0.64-1.8~ {\rm \mu m}$), and high spectral resolution ($R\sim4000-7000$) will enable SDSS-like surveys at high redshifts for the first time, revolutionising our understanding of galaxy mass assembly and chemical evolution around cosmic noon (\citealp{Cirasuolo2020}).
In particular, the planned guaranteed time observation (GTO) survey MOONS Redshift-Intensive Survey Experiment (MOONRISE; \citealp{Maiolino2020}), utilising hundreds of nights of guaranteed time, aims to obtain high-quality spectra for up to half a million galaxies at $0.9 < z < 2.6$. This unprecedented sample will allow for robust measurements of key physical properties including metallicity, SFRs, AGN activity, and stellar populations across a wide range of environments and galaxy masses. By accessing the same rest-frame optical diagnostics used in statistically significant studies of the local universe, MOONS will offer a consistent probe of galaxy evolution over cosmic time, critically constraining models of galaxy formation and transformation.
We highlight the efficacy of \texttt{M-TOPnet} in accurately predicting the redshift, $M_{\rm star}$, and SFR from simulated MOONS spectra of galaxies at $z\sim1-3$, even for faint sources for which traditional methods often struggle. 
Our model proves to be a practical and adaptable approach to tackle the upcoming challenges in observational astronomy. By leveraging multi-task learning and the capability to handle uncertainties by outputting probabilities (see Section \ref{sec:methods}), it delivers the precision, the efficiency, and the flexibility needed to robustly analyse the massive, complex datasets expected from next-generation instruments.
\\\\
This paper is organised as follows. 
Section \ref{sec:dataset} describes the origin and properties of the simulated MOONS dataset. 
In Section \ref{sec:methods}, we outline our methodologies, covering data processing, preparation, and the design and training of \texttt{M-TOPnet}. 
Section \ref{sec:results} presents the results, with further elaboration provided in the discussion (Section \ref{sec:discussion}). Finally, Section \ref{sec:conclusions} provides the conclusions.

\section{The simulated MOONS dataset}
\label{sec:dataset}

\begin{figure*}[ht!]
\centering
\includegraphics[width=1\textwidth]{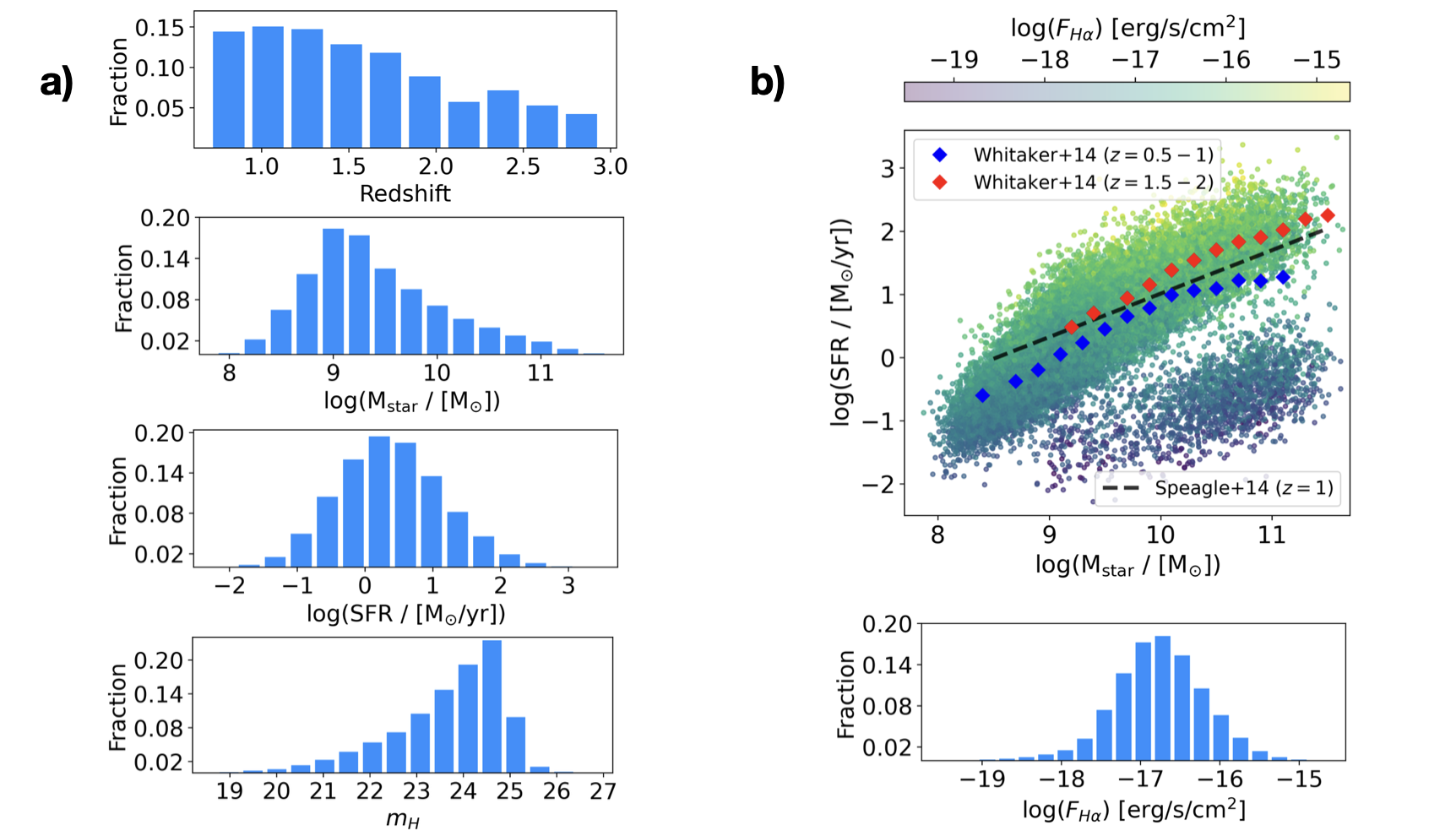}
\caption{Properties of galaxies in the simulated dataset. 
{\bf (a)} Distributions of key galaxy properties, including redshift (first panel), $M_{\rm star}$ (second panel), SFR (third panel), and $m_H$ (bottom panel), 
arranged from top to bottom.
{\bf (b)}
Upper panel: Distribution of galaxies in the log($M_{\rm star}$) versus log(SFR) diagram, highlighting their representation of both main-sequence star-forming galaxies and passive galaxies. 
Galaxies are colour-coded by their $\rm H\alpha$ line flux. 
For reference, we show the trends of the main sequence at $z=0.5-1$ (blue points) and at $z=1.5-2$ (red points) reported by \cite{Whitaker2014}, and the best fit for the main sequence at $z\sim1$ by \cite{Speagle2014}.
Lower panel: flux distribution of the $\rm H\alpha$ line.
}
\label{fig:dataset_1}
\end{figure*}

\begin{figure*}[ht!]
\centering
\includegraphics[width=1\textwidth]{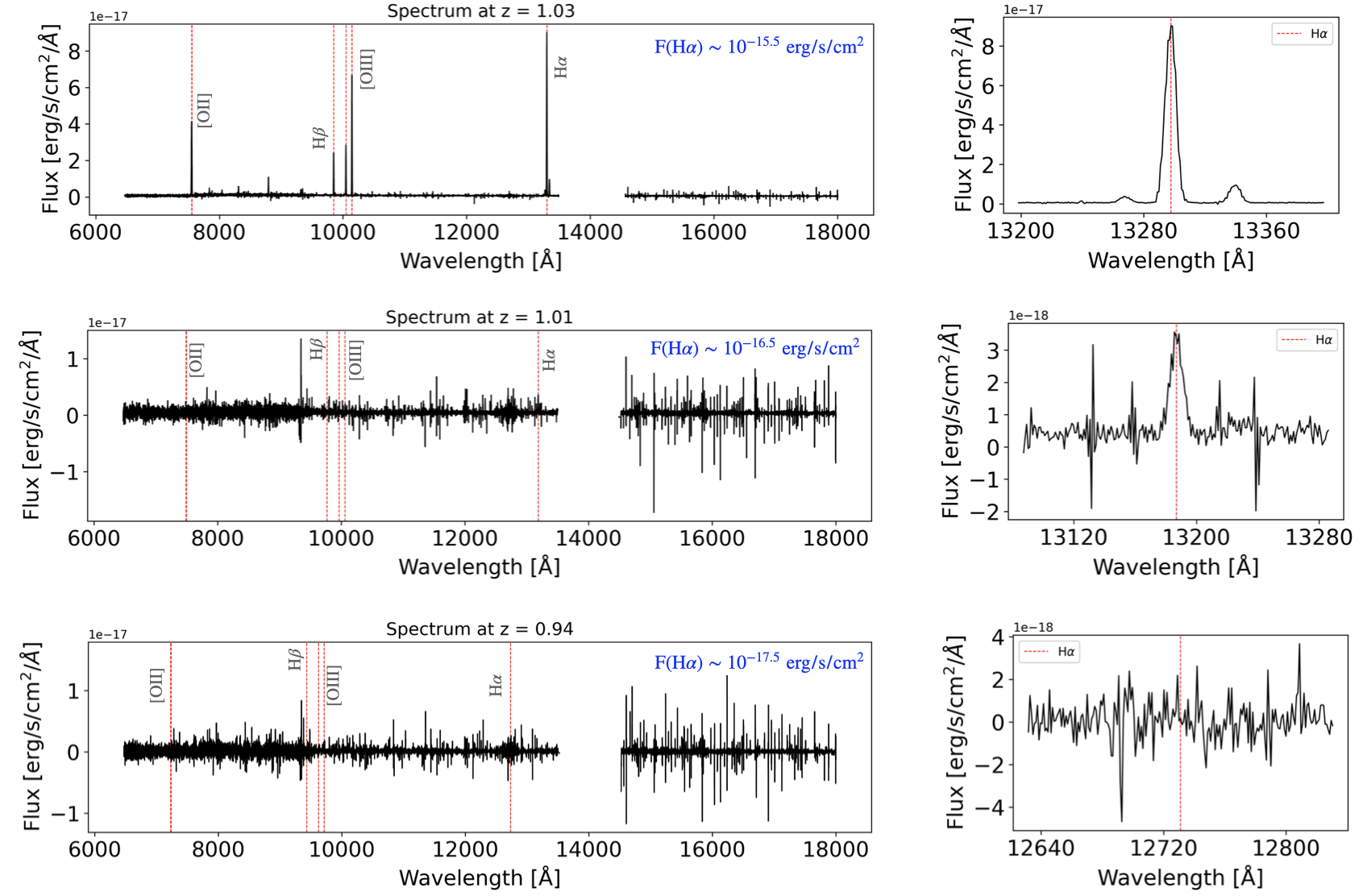}
\caption{
Examples of simulated galaxy spectra from our sample, shown before any pre-processing required for input into our pipeline, covering the entire spectral range.
Each row represents a different galaxy spectrum, with the left panels showing the full spectrum and the right panels providing a zoomed-in view around the $\rm H\alpha$ line. The spectra are positioned based on the $\rm H\alpha$ line flux: $\rm F(H{\alpha}) \sim 10^{-15}~erg/s/cm^2$ (upper row), $\rm F(H{\alpha}) \sim 10^{-16.5}~erg/s/cm^2$ (middle row), and $\rm F(H{\alpha}) \sim 10^{-17.5}~erg/s/cm^2$ (lower row), representing high-, average-, and low-quality spectra, respectively. Red vertical lines indicate the positions of key emission lines, including [OII], H$\beta$, [OIII], and H$\alpha$, with corresponding labels. The missing flux values around 14000 $\AA$ are due to the gap between spectral bands (see Section \ref{sec:dataset}).}
\label{fig:dataset_2}
\end{figure*}

In this work, we employ a dataset of simulated MOONS spectra to train and validate our 
\texttt{M-TOPnet} pipeline, in preparation of future real observations.
The input mock spectra are generated by {\sc MAMBO} (Mocks with Abundance Matching in BOlogna), a method devised to quickly paint galaxy properties ($M_{\rm star}$, SFR, photometry, dust, metallicity, size, decomposition bulge or disc, emission lines, and spectra) on top of Dark Matter (DM) sub-haloes and create realistic multi-wavelength mock data for the next generation of surveys (see a discussion on {\sc MAMBO} in \citealp{2024arXiv240906700L}).  
In the present work, we run the {\sc MAMBO} workflow on a lightcone built by \cite{Henriques2015} on the Millennium Simulation (\citealp{Springel2005}), namely lightcone number 23, which has a mass function closest to the mean of all 25 available lightcones. 
The lightcone covers the redshift range $0<z<10$ and contains DM haloes with $M_{\rm sub} > 1.7 \times 10^{10}\,M_\odot\,h^{-1}$, with an area of $3.14$\,deg$^2$. 

To assign properties to galaxies, MAMBO first assigns the $M_{\rm star}$ using the Stellar-to-Halo Mass relation \citep{Girelli2020} derived from the observed Stellar Mass Function (SMF) at $z \sim 0$ from SDSS \citep{Peng2010}, the SMF in COSMOS for redshifts between $0.2 < z < 4$ \citep{Ilbert2013}, and the SMF from CANDELS for $z \geq 4$ \citep{Grazian2015}. 
Each galaxy is then classified as either quiescent or star-forming, with a small random fraction designated as starbursts, based on the relative proportions of SMFs for the red and blue populations. This assignment ensures that more massive, predominantly quiescent galaxies are linked to large sub-halos, typically located in the dense, central regions of massive halos, thus accurately reflecting the observed correlation between galaxy colour and environment (\citealp{2024arXiv240906700L}).

All other galaxy properties are determined using a modified version of the open-source code {\sc EGG} \citep[Empirical Galaxy Generator;][]{Schreiber2017}\footnote{\url{https://cschreib.github.io/egg/}}, which has been thoroughly validated against extensive observations and distributions of physical properties (for a detailed description, see \citealp{Girelliphd}, and \citealp{2024arXiv240906700L} for the AGN modelling). {\sc EGG} applies observed scaling relations, such as the star-forming main sequence \citep{Schreiber2015} to estimate SFR, distinguishing between UV-derived obscured SFR and IR-derived dust-free SFR. 
The H$\alpha$ luminosity is inferred from the Kennicutt-Schmidt law \citep{Kennicutt1998a}, and the  hydrogen lines from line ratios for Case B recombination and $T=10000$\,K as in \cite{Osterbrock2006}. 
The other emission lines are generated using stellar masses and SFRs, and calibrated on the SDSS, assuming redshift evolution as in \cite{Shapley2015}.
Gaussian emission lines are simulated with a velocity dispersion with a mass-dependent $\sigma_{\rm gas}$ from \citet{Bezanson2018}.

The code also generates rest-frame and observed photometry from UV to submillimetre wavelengths, based on the SEDs of both bulge and disc components and using specific broad-band filters for MOONRISE \citep{Maiolino2020}. At optical and near-IR wavelengths, the SEDs are drawn from a library of templates built from the \cite{Bruzual2003} models, dust-attenuated using Calzetti's law \citep{Calzetti2000}, and cover the $UVJ$ colour space \citep{Williams2009} for both quiescent and star-forming galaxies.
The attenuation of emission lines is differential compared to continuum, considering the redshift dependence derived in \cite{Pannella2015}.
From the full galaxy sample, we extracted a sub-sample that mimics the MOONSRISE selection. Specifically, we selected sources with apparent magnitude in the H band $m_H < 25$ 
or both log($M_{\rm star}/M_\odot$)$>$9 in the redshift range $0.7<z<3$, and covering an area close to a single MOONS field of view. 
In total we selected about 40 000 galaxies, for which MAMBO creates also the observed spectra of the stellar continuum at vacuum wavelengths, to which we added emission lines according to catalogue fluxes, applied the assigned velocity dispersion and resampled according to MOONS pixel. We, therefore, assigned a size to each galaxy according to their mass using \cite{VanderWel2014} relations for passive and star forming, separately. We considered the light loss due to the MOONS fibre using Sérsic profiles, assuming Sérsic index $n=4$ and $n=1$, for passive and star-forming galaxies, respectively. 
\\

The final mock spectra used for this work are generated by running MAMBO templates through {\sc moons1d}. 
The {\sc moons1d} package
\footnote{\url{https://gitlab.physics.ox.ac.uk/rodriguesm/moons1d}}
is a {\it Python}-based 1D spectral simulator specifically designed by the MOONS collaboration for the MOONS instrument at the VLT. This tool generates simulated spectra based on observing conditions and input templates, including options for different observing bands (such as RI, YJ, and H) resolutions, and observing strategies (XSWITCH and STARE; see \citealp{Maiolino2020}). Users can specify parameters like seeing, airmass, and exposure time to simulate realistic observational conditions. The output is a FITS file containing the simulated spectrum. {\sc moons1d} has been instrumental in testing the science pipelines, supporting the preparation for future observations by providing realistic mock spectra.

For the dataset used in this work, for any given {\sc MAMBO} template, {\sc moons1d} is ran using the low-resolution mode for all three MOONS spectral channels (RI, YJ, H), adopting a range of 2, 4 and 8 $h$ of exposure time (assuming individual exposures of 300 $s$).
Also, a seeing of 0.8$^{''}$ and an airmass of 1.2 are assumed, and the spectra include the effects of fibre loss. We have also adopted a simulated XSWITCH observing strategy (see \citealp{Maiolino2020}), i.e, a technique for high-efficiency nodding observations. 
In this approach, two fibers are allocated per object: one fiber captures the target’s signal, while the other is positioned on a sky region. During the observation, the target and sky fibers alternate positions, allowing one to collect data from the target and the sky in the same fiber, which significantly improves sky subtraction accuracy.
This strategy is planned to be adopted in the MOONRISE survey (\citealp{Cirasuolo2020}) as it will be particularly valuable for faint object observations.
\\

As a result, we obtain a mock dataset of 118.194 spectra, each composed by 12.217 spectral elements, covering a wavelength domain (over the three RI, YJ, H spectral bands) from 0.64 to 1.8 $\rm \mu m$, with a gap between about 1.35 and 1.45 $\rm \mu m$. 
The redshift, $M_{\rm star}$, and SFR of each spectrum -- that is, the physical properties that we aim to derive in this work -- are taken as ground truths from {\sc MAMBO}.

The distributions of the key quantities of galaxies in the simulated dataset are illustrated in Figure~\ref{fig:dataset_1}a.
In detail, we show histograms of redshift, $M_{\rm star}$, SFR and $m_H$.
Redshift values range from approximately 0.7 to 2.9, although their distribution is not uniform and tend to peak towards the low-edge tail.
Values of $M_{\rm star}$ are distributed between about $10^{8}~{\rm M_{\odot}}$ and $10^{11.2}~{\rm M_{\odot}}$, with a peak around $10^{9}~{\rm M_{\odot}}$, while SFRs range between $10^{-2}~{\rm M_{\odot} / yr}$ and $10^{3}~{\rm M_{\odot} / yr}$, with a peak at about $2-3 ~{\rm M_{\odot} / yr}$.
Values of $m_H$ are mostly distributed between about 19 and 27, with $m_H \sim 24-25$ galaxies being the most represented. 
In the upper panel of Figure~\ref{fig:dataset_1}b, we report the distribution of our galaxies in the log($M_{\rm star}$) versus log(SFR) diagram, 
showing that galaxies in our sample are representative of both populations of normal main-sequence star-forming galaxies and passive galaxies. 
Galaxies in this diagram are colour-coded according to their $\rm H\alpha$ line flux, whose distribution is shown in the lower panel of Figure~\ref{fig:dataset_1}b.

In Figure~\ref{fig:dataset_2}, we show three examples of simulated galaxy spectra in our sample, covering the entire spectral domain. 
The right panels provide a closer look at the zoomed spectrum around the $\rm H\alpha$ line.
With $\rm H\alpha$ line fluxes of $\rm F(H{\alpha})\sim 10^{-15} ~erg/s/cm^2$ (upper panel), $\rm F(H{\alpha})\sim 10^{-16.5} ~erg/s/cm^2$ (central panel), and $\rm F(H{\alpha})\sim 10^{-17.5} ~erg/s/cm^2$ (lower panels), these spectra can be considered representative examples
at $z\sim1$
of high-, average-, and low-quality spectra in our sample, respectively.   
These spectra were generated with a simulated exposure time of $2~h$, resulting in a fixed noise level. Notably, the third spectrum represents a quiescent galaxy, indicating that such spectra are particularly challenging for this study.

\section{Methods}
\label{sec:methods}

\subsection{Data processing and preparation before training}
\label{sec:preprocess}

After combining the RI, YJ, and H spectral bands (see Section \ref{sec:dataset}), we performed several preprocessing steps on the spectra before training the \texttt{M-TOPnet} pipeline outlined in the subsequent sections. These steps were performed on the entire dataset (prior to splitting it for training, validation, and testing) and are intended to be applied to any new instances for future inference once the \texttt{M-TOPnet} pipeline is trained.

Initially, we applied sky masking to our spectra to mitigate the impact of strong sky lines, assuming a sky model provided by the simulations. This process involves masking spectral channels exceeding a conservative flux threshold of $10^{-4.8} ~ {\rm photons/s/cm^2/\AA/arcsec^2}$ in the pre-defined sky model, effectively ignoring channels where sky emission might dominate. This threshold corresponds to rejecting approximately 15\% of the channels across our spectra. Variations in this threshold have been tested, demonstrating the robustness of our results with up to 25\% channel exclusion.

Then, we removed the continuum from the spectra using a running median filtering technique, subtracting a median-filtered spectrum with a window size of 1000 channels from the original. This size was chosen to effectively capture the overall shape of the continuum emission. While more precise methods such as full spectral fitting exist (see Section \ref{sec:intro}, and references therein), they tend to be slower; our approach provides a practical balance of speed and accuracy. 
Subsequently, we normalised the continuum-subtracted spectra by dividing them by their maximum value in each spectrum. This two-step procedure ensures that the resulting spectra are free of continuum contribution and normalised, facilitating further analysis of spectral lines.

The continuum, estimated through running median filtering, was then rebinned into ten points using linear interpolation. This reduction in dimensionality of the continuum preserves essential shape characteristics, and complements the information contained in the continuum-subtracted and normalised spectra. 
These continuum points are used alongside the emission lines to derive physical properties of galaxies, providing a more comprehensive analysis of the spectral data. Our approach effectively captures the underlying continuum variations while simplifying the input features for subsequent modelling.
As a result, for each spectrum, we generate a normalised continuum-subtracted spectrum to emphasise the spectral lines, and a ten-point vector encoding information on the continuum shape, which is not yet normalised but will be standardised according to the distribution of the training set, as is explained in the following paragraphs. We note that this modular design allows for future enhancements, such as incorporating photometric points from imaging into the continuum vector.

At this point, our dataset consists of 1) normalised, sky-masked, and continuum-subtracted spectra, 2) continuum vectors, and 3)  galaxy property labels that we aim to predict; that is, redshift, $M_{\rm star}$, and SFR. 
To train and test \texttt{M-TOPnet}, we split the dataset into training, validation, and test sets. The training set accounts for 70\% (82.736 examples) of the sample, while both the validation and test sets account for 15\% each (17.729 spectra). We note that spectral and galaxy properties of instances in training, validation, and test sets follow the same distributions as are shown in Figure~\ref{fig:dataset_1}a and are thus representative of the total simulated MOONS dataset.
After splitting, prior to training, the continuum vectors and labels for the logarithms of $M_{\rm star}$ and SFR were  
standardised separately, based on the mean and standard deviation values from the training set. This standardisation was applied consistently across the training, validation, and test sets. By doing so, we preserve the relationships between the continuum and the physical properties like $M_{\rm star}$ and SFR, while ensuring data consistency across the dynamic range, facilitating model training.

We do not need to normalise the redshift as it is not shown in training as a scalar value. 
Instead, as is explained in the following and anticipated in Section \ref{sec:intro}, we approached redshift prediction as a classification task by discretising continuous redshift values into finely spaced bins. 
We defined these bins with a resolution of $dz = 0.003$, optimising the balance between granularity and the limitations given by the spectral resolution. This choice corresponds to velocity bins ranging from approximately 200 to 500 $\rm km ~s^{-1}$ at $z\sim0.7-3$, which is somewhat broader than the $< 100 ~\rm km ~s^{-1}$ resolution expected for MOONS in low-resolution mode. However, we find it to be the optimal compromise between achieving sufficient accuracy for robust redshift determination and maintaining relatively compact discretised vectors of redshift elements, allowing us to conduct multiple tests on the pipeline.
We note that similar results to those discussed in Section \ref{sec:results} were obtained in a single test using a finer bin width of $dz = 0.0005$, which is more consistent with the nominal instrumental resolution. However, this finer discretisation increased the vector size to nearly 5000 elements, significantly slowing down both training and inference processes, and making it infeasible to perform all the tests needed to fully characterise the pipeline and identify the best model. 
Therefore, we have opted for $dz = 0.003$ in this version of the pipeline, leaving further refinements for future work.

To represent redshift input values, we employed a modified one-hot vector representation, where each redshift is represented by a Gaussian distribution centred on its corresponding bin. 
This method allows us to adapt to the inherent uncertainty in real observational data, where lower-quality spectra may yield less precise redshift measurements. 
Typically, this is crucial for real data, where spectral quality can affect redshift accuracy. 
However, since our analysis uses simulated data with no error associated to the labels, we employed a standard Gaussian, setting its standard deviation $\sigma$ to a fixed value of 0.001. 
We also tested with a $\sigma$ of zero-channels (i.e. a Dirac Delta function) or by scaling the sigma linearly with the standard deviation of the continuum-subtracted spectra, but we did not see any significant change in the results. We note that these redshift PDFs are already normalised by definition, so they do not need further normalisation.

A final step in data preparation consisted of generating an auxiliary spectrum for each galaxy spectrum, which serves as a training tool for line-location tasks within our multi-task neural network (see a discussion in the following Section \ref{sec:neuralnet}). 
This auxiliary spectrum was populated with zeros, except at positions corresponding to expected emission and absorption lines based on the redshift. 
To account for spectral dispersion and minor offsets in line positions, and thus help the identification of the line location, we also marked five adjacent channels on either side of each expected line with ones. This parameter was tested across a range from 1 to 10 adjacent channels, with negligible impact on results. The emission and absorption lines used to define the spectral line masks are reported in Table \ref{tab:spectral_lines}.

\begin{table}

\centering
\caption{Spectral lines (both in emission and in absorption), and their rest-frame vacuum wavelengths, used to generate the auxiliary spectra containing spectral line masks, as is discussed in Section \ref{sec:preprocess}. 
}
\begin{tabular}{ll}
\hline
Spectral Line & Rest-frame Wavelength (\AA) \\
\hline
Ly$\alpha$ & 1215.2 \\
N V & 1240.8 \\
Si II & 1260.4 \\ 
O I & 1305.5 \\ 
C II & 1335.3 \\
Si IV & 1393.8, 1402.8 \\ 
Si II & 1526.7 \\
C IV & 1549.5 \\
He II & 1640.4 \\
{[}O III{]} & 1660.8, 1666.2 \\ 
C III{]} & 1908.7 \\ 
{[}O II{]} & 3727.1, 3729.9 \\ 
Ca II H\&K & 3933.7, 3968.5 \\ 
CN & 3875 \\ 
H$\beta$ & 4862.7 \\
{[}O III{]} & 4960.3, 5008.2 \\
Fe I & 4669.3 \\ 
Mg I & 5167, 5172, 5183 \\ 
Fe I & 5270 \\ 
Na I D & 5892 \\ 
TiO & 6180 \\ 
{[}N II{]} & 6549.9 \\
H$\alpha$ & 6564.6 \\
{[}N II{]} & 6585.3 \\
{[}S II{]} & 6718.3, 6732.7 \\
TiO & 7150 \\ 
Na I & 8183, 8195 \\ 
Ca II & 8489, 8542, 8662 \\ 
{[}S III{]} & 9070, 9532 \\ 
He I & 10830.3 \\ 
\hline
\end{tabular}
\label{tab:spectral_lines}
\end{table}

\subsection{Neural network design and training}
\label{sec:neuralnet}

\begin{figure*}[ht!]
\centering
\includegraphics[width=1\textwidth]{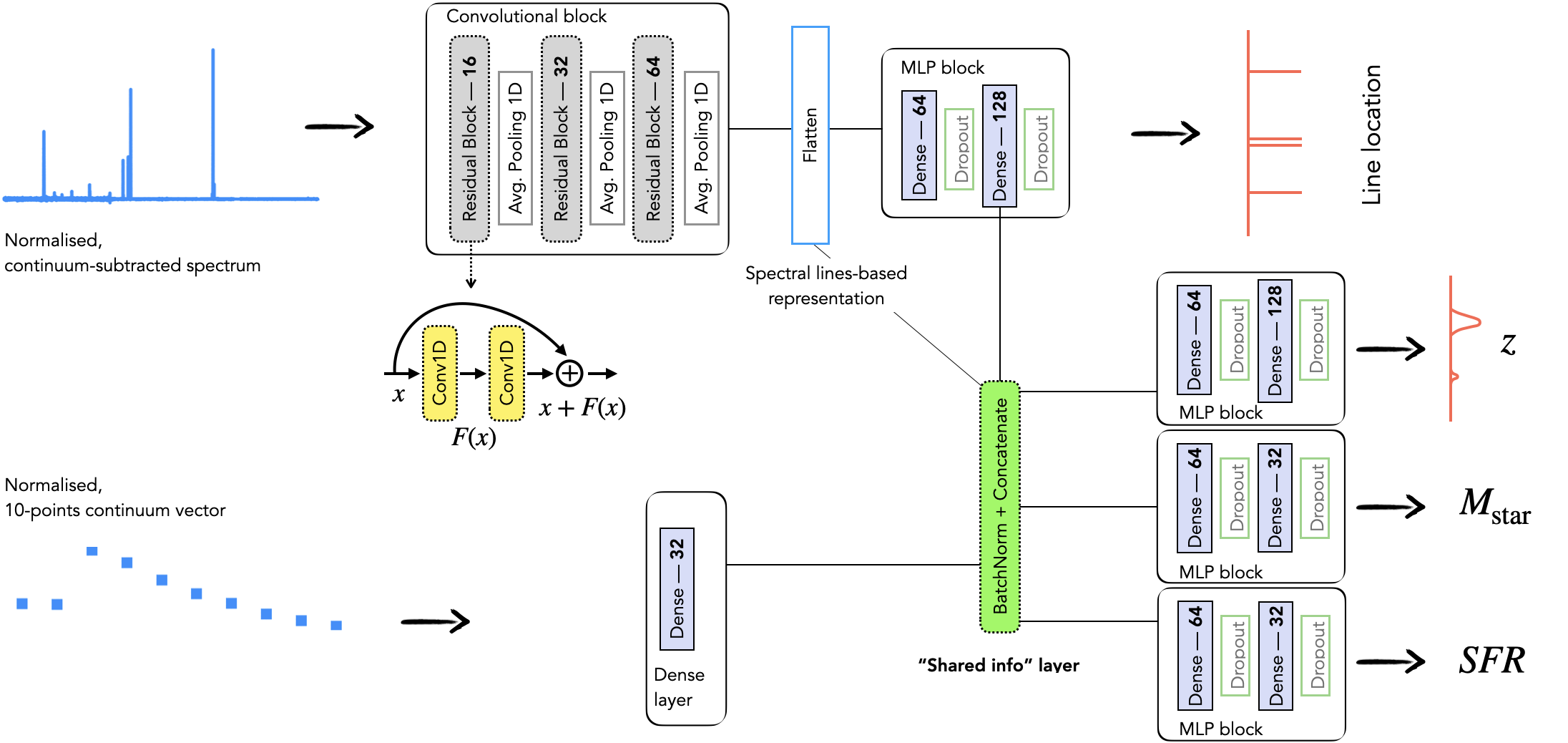}
\caption{Architecture of \texttt{M-TOPnet}, the deep learning pipeline used for the galaxy spectral analysis in this work. Our model accepts two inputs: a normalised, continuum-subtracted spectrum and a normalised continuum vector. The spectrum is processed through convolutional residual blocks to extract spectral features, while the continuum vector is encoded through a dense layer. Outputs from both branches, along with intermediate outputs from the line location task, are combined into a shared information layer. This shared layer feeds into task-specific branches consisting of MLPs to predict line locations, redshift, and normalised values of $M_{\rm star}$, and SFR. See a discussion on the different components in Section \ref{sec:neuralnet}.} 
\label{fig:nn-sketch}
\end{figure*}

We designed a multi-task neural network model to simultaneously predict redshift, $M_{\rm star}$, SFR, and perform spectral line location. 
The architecture has been developed using \texttt{Keras}, a high-level neural network API that runs on top of \texttt{TensorFlow} (\citealp{chollet2015keras, Abadi2016}). 
The model architecture is tailored to process both the continuum-subtracted spectra and the ten-point continuum vectors, leveraging the strengths of convolutional layers.
CNNs have been widely utilised in ML (\citealp{Goodfellow2016}), demonstrating substantial improvements in computer vision, and have become widely adopted in astronomy (see Section \ref{sec:intro} and references therein).
While their usage has been mostly focussed on the image domain 
(e.g. \citealp{Dieleman2015, Zhu2019, Wu2019, Smith2023}),
CNNs are also particularly useful for galaxy spectra because they can capture hierarchical features at different scales. The convolution operations detect local patterns, while pooling operations help capture wider-scale dependencies, making them well suited for analysing spectral features that exist across various scales (see e.g. \citealp{Stivaktakis2018, Melchior2023, Wu2023, Wang2024}).
The pipeline also makes use of multi-layer perceptrons (MLPs), which are fully connected neural networks consisting of multiple layers of neurons, each applying a non-linear activation function to its inputs.

\subsubsection{Model architecture}

Here, we discuss the architecture of our \texttt{M-TOPnet} model, sketched in Figure \ref{fig:nn-sketch}, outlining its fundamental parts.

\begin{itemize}

\item {Input layer --} The model accepts two inputs: continuum-subtracted spectra and continuum points. These are processed separately due to their distinct nature and require transformations. \newline

\item {Spectral processing --} The continuum-subtracted spectra undergo a series of convolutional residual blocks. Residual blocks, introduced by \cite{He2015}, allow for the training of deeper networks by addressing the vanishing gradient problem. 
They achieve this by introducing skip connections that bypass one or more layers, enabling the network to learn residual functions with reference to the layer inputs (see an application to galaxy spectra in \citealp{Moradi2024}). In \texttt{M-TOPnet}, each residual block consists of two convolutional layers followed by an average pooling layer with a pool size of five, which means that each pooling operation reduces the spatial dimensions by a factor of five. 
The number of filters increases from 16 to 64 across the blocks. 
This approach enables the model to capture simple features in initial layers and progressively more complex patterns in deeper layers, exploiting the hierarchical nature of spectral data and improving its ability to represent intricate spectral characteristics. This strategy has been effectively employed in other cases using astronomical spectra (see e.g. \citealp{Podsztavek2022, Ambrosch2023, Wang2024}), and we found it to be the most efficient architecture for our tasks.
The convolutional layers use the rectified linear unit (ReLU) activation function, which is widely used due to its ability to mitigate the vanishing gradient problem and accelerate convergence (\citealp{Nair2010}). \newline

\item {Continuum processing --} The continuum points are processed through a dense layer, which consists of 32 neurons,
converting the low-dimensional continuum information into a higher-dimensional representation. \newline

\item {Feature combination --} The flattened output from the spectral processing branch, the encoded continuum information, and the intermediate output from the line location task are concatenated to form a `shared info' layer. This layer is preceded by a batch normalisation layer, which helps to stabilise the learning process and reduce internal covariate shift. 
The `shared info' layer efficiently encodes the information of the spectra and of the continuum. In Section \ref{sec:discussion}, we show through a dimensionality reduced visualisation that this layer effectively learns a general representation of the spectra from which one can retrieve physical information that was not used as labels during training. \newline

\item {Task-specific branches --}
Each task-specific branch consists of an MLP with dropout layers, as follows.
\begin{itemize}
    \item[a.] Line location: an MLP with two hidden layers (64 and 128 neurons) and sigmoid activation in the output layer predicts the locations of spectral lines. We use sigmoid activation because each spectral channel in the input can be 0 or 1, making it a binary classification problem.
    \item[b.] Redshift: an MLP with two hidden layers (64 and 128 neurons) and Softmax activation in the output layer performs redshift classification, outputting a probability distribution over the redshift bins. This discrete approach via Softmax provides an efficient solution for our classification task within the discretised redshift space. 
    \item[c.] Stellar mass: an MLP with two hidden layers (64 and 32 neurons) and linear activation in the output layer predicts log($M_{\rm star}$).
    \item[d.] Star formation rate: an MLP with two hidden layers (64 and 32 neurons) and linear activation in the output layer predicts log(SFR).
\end{itemize}
We note that both predictions for log($M_{\rm star}$) and SFR are standardised and need to be scaled back to their original dynamic range using the inverse transformation of the standardisation before any analysis.
Each hidden layer in these MLPs is followed by a dropout layer with a dropout rate of 0.2. 
\end{itemize}

Dropout is a regularisation technique introduced that helps prevent overfitting by randomly dropping out a fraction of neurons during training (\citealp{Hinton2012,Srivastava2014}). 
This process helps to reduce the network's reliance on any particular set of features. This leads to better generalisation and reduces overfitting. The technique has been shown to be particularly effective in deep neural networks and has become a standard tool in deep learning pipelines.
In this work, we have experimented with keeping dropout switched on during inference as well, which is a technique known as MC dropout (\citealp{Gal2015}). This approach allows us to estimate the model's epistemic uncertainty by performing multiple forward passes through the network with different dropout masks (see Section \ref{sec:intro} and Section \ref{sec:results}).

\subsubsection{Multi-task learning}
The multi-task approach is particularly beneficial in our context. 
In general, multi-task learning has been proven to have several advantages (see e.g. \citealp{Caruana1997, Ruder2017, Crawshaw2020}): it allows for parameter sharing across tasks, which can lead to better generalisation; it acts as a regulariser, reducing the risk of overfitting; it can leverage the commonalities and differences across tasks to improve learning efficiency; and it can potentially lead to better performance on all tasks compared to single-task models. 

While our primary focus is on accurate redshift determination, we found that including the line location task significantly improves the overall performance, especially for redshift prediction. This auxiliary task helps the model develop a better internal representation of the spectral features, even though it may not always perform optimally on its own (as is discussed in Section \ref{sec:results}). 
The inclusion of $M_{\rm star}$ and SFR prediction tasks further enriches the shared representations, capturing more detailed spectral characteristics that indirectly benefit the redshift determination. 
This method leverages the existing physical relationships between the continuum (in terms of both its normalisation and shape) and spectral lines relative to $M_{\rm star}$ and SFR, which, in turn, enhances the model's ability to infer redshifts more effectively.
We note that the multi-task approach is innovative because it involves strategically defining related auxiliary tasks (such as line location, in our case), which impose implicit constraints on the model. 
These tasks help the model learn physically relevant features, ultimately guiding it towards a more accurate redshift estimation.

It is worth mentioning that multi-task learning should not be confused with multi-output regression. 
Unlike multi-output regression, which typically handles multiple  outputs with a single type of loss function, multi-task learning can integrate tasks of different natures with task-specific loss functions, as is demonstrated in our joint tasks of probability distribution estimation for the redshift, continuous regression for $M_{\rm star}$ and SFR, and discrete line location.
Multi-task learning thus provides added flexibility by allowing each task to have its own objective function, enabling the model to learn from distinct yet complementary tasks, which can enhance performance and generalisation. We discuss the impact of multi-task learning on our pipeline in Section \ref{sec:results}.

\begin{figure*}[ht!]
\centering
\includegraphics[width=1\textwidth]{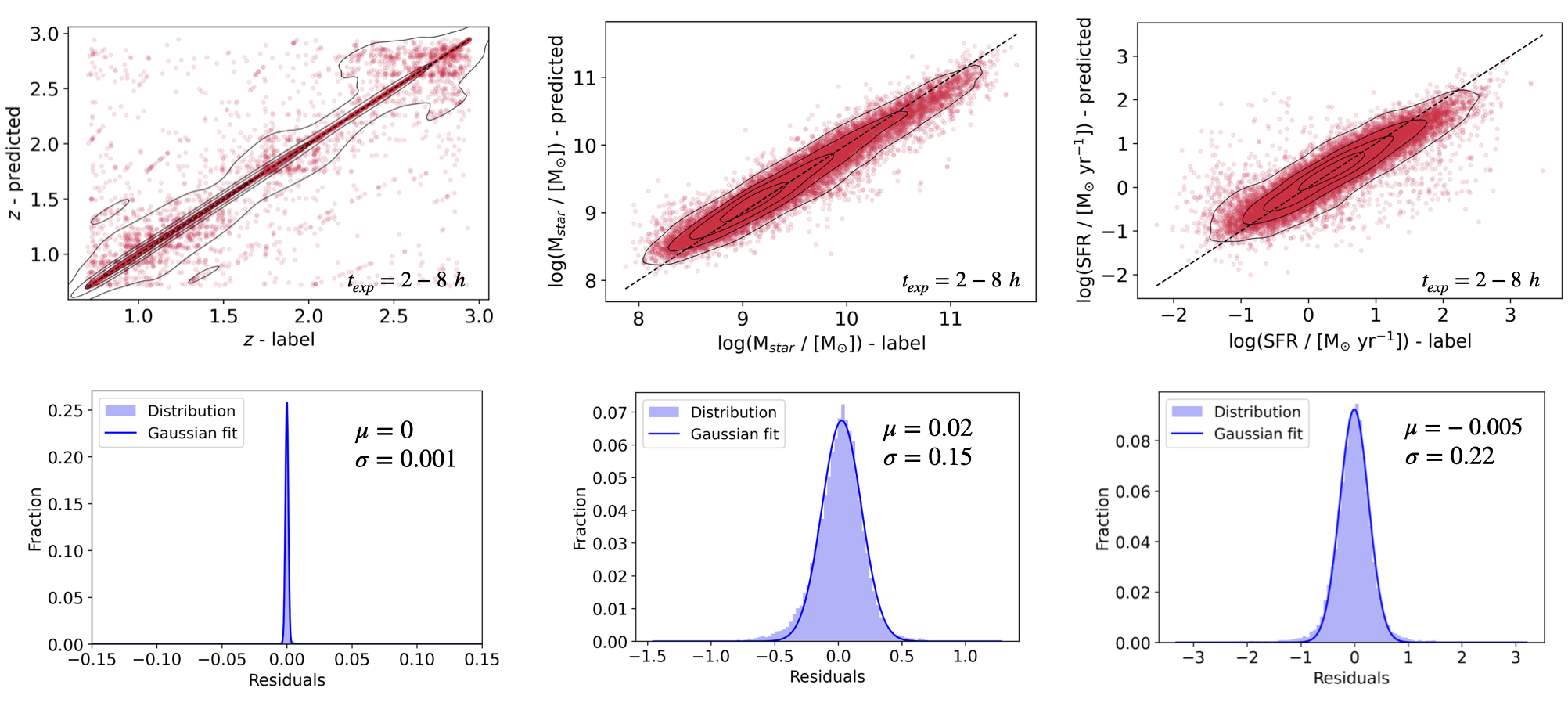}
\caption{Comparison of predicted and labelled values for redshift, $M_{\rm star}$, and SFR for all simulated MOONS galaxy spectra in the test set with exposure times of $2~h$, $4~h$, and $8~h$. Upper panels: 
Scatter plots of predicted versus labelled values, with the dashed black line indicating perfect agreement. Solid black lines represent the contours of the joint probability density, corresponding to the 20th, 50th, 75th, and 95th percentiles of the distribution, estimated using a Gaussian kernel density estimator. 
Lower panels: Histograms of the residuals (‘predicted value - labelled value’; see Section \ref{sec:results}) with Gaussian fits, showing the mean ($\mu$) and standard deviation ($\sigma$) of the distributions.}
\label{fig:residuals1}
\end{figure*}

\subsubsection{Training process}

The model was trained using a combined loss function, weighing the contributions from each task (see below). We employed an adaptive learning rate strategy, starting with an initial rate of 0.005 and decreasing it linearly to 0.0005 over 50 epochs, after which it remains constant. 
This approach follows fairly standard practice (\citealp{Goodfellow2016, Defazio2023}), where the initial and final learning rates are chosen to allow for a balanced combination of rapid convergence and stable refinement.
The training process was set to run for a maximum of 100 epochs with early stopping based on validation loss, using a patience of 10 epochs. This approach helps prevent overfitting while allowing the model sufficient time to converge. 
Our best model converged after 60 epochs, although convergence generally oscillates by a factor of $\pm 8$ epochs, likely due to weight initialisation and data shuffling in the mini-batches. 

The loss functions for each task are as follows: categorical cross-entropy for redshift, binary cross-entropy for line location, and mean squared error for stellar mass and SFR prediction. We assigned different weights to each task's loss, with the emission line location task given a higher weight (10) compared to the others (1 each). 
This weighting scheme was determined to be optimal through experimentation. We also tested equal weighting across all tasks and other variations, but found that emphasising the line location task led to the best overall model performance.
This approach makes intuitive sense: if the network cannot accurately identify emission lines, it will struggle to predict the redshift, whereas predictions for $M_{\rm star}$ and SFR are less sensitive to precise line identification.
\\\\
In conclusion, our \texttt{M-TOPnet} architecture leverages the strengths of both CNNs and MLPs, utilising residual blocks for efficient spectral processing and a multi-task learning approach to improve overall performance, particularly in redshift determination. The model design allows it to extract and combine information from both the spectral lines and the continuum, providing a comprehensive analysis of the input data. We shall make the code for our pipeline, including the \texttt{M-TOPnet} architecture, publicly available on GitHub after the paper is accepted.

\section{Results}
\label{sec:results}

\begin{figure*}[ht!]
\centering
\includegraphics[width=0.9\textwidth]{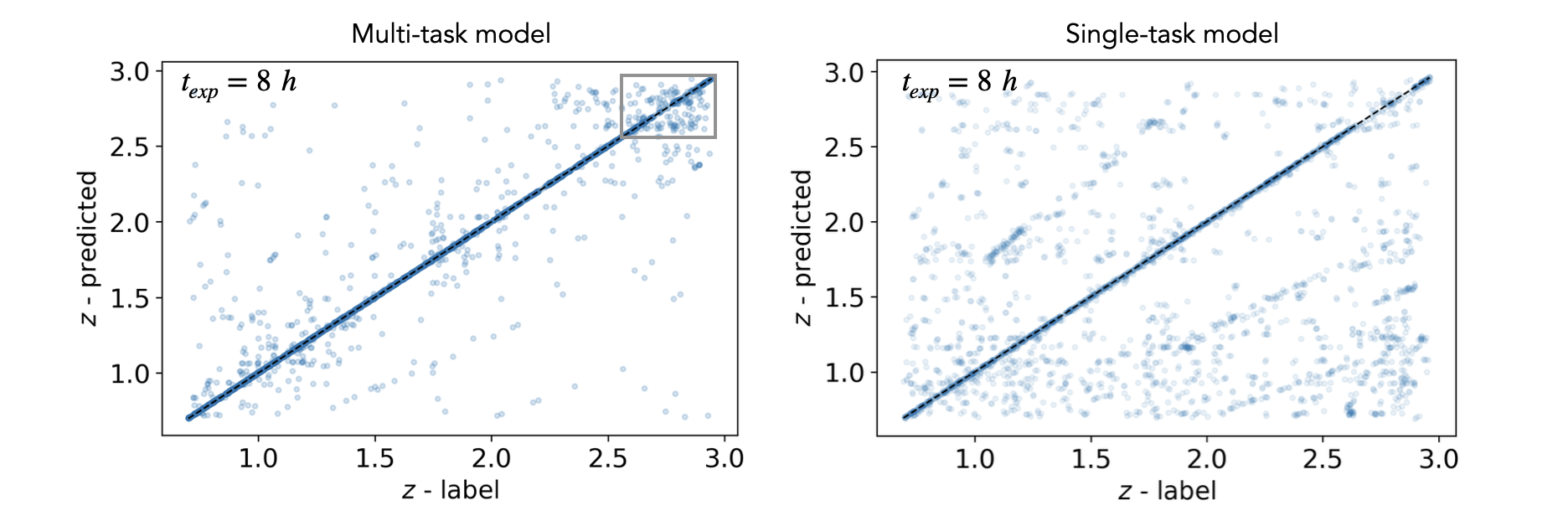}
\caption{Comparison of redshift prediction accuracy for 8-hour spectra using two models: the multi-task model discussed in Section \ref{sec:methods} on the left panel, and a single-task model (showing increased outliers, including catastrophic errors; see Section \ref{sec:results}) on the right panel.
The redshift range $z = 2.7 - 3$, where a significant relative excess of unsuccessful predictions is shown (see Section \ref{sec:discussion} for a discussion) is highlighted by the grey box.
}
\label{fig:residuals2}
\end{figure*}

\begin{figure*}[ht!]
\centering
\includegraphics[width=1\textwidth]{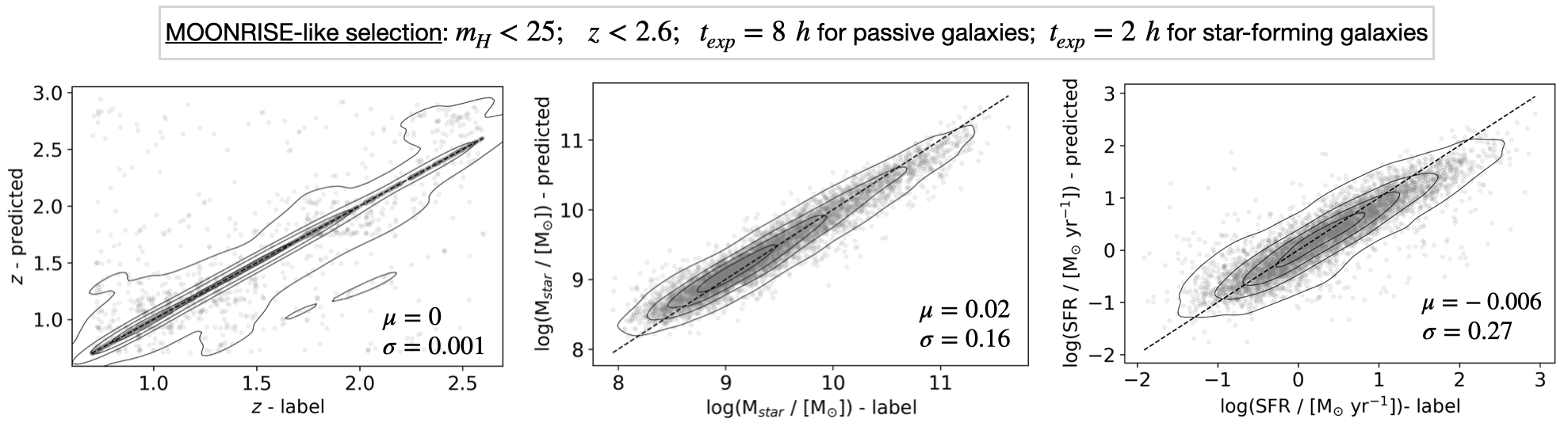}
\caption{Same as the upper panels of Figure \ref{fig:residuals1}, but indicative of the subset of galaxies defined by  selection criteria similar to the planned MOONS GTO survey MOONRISE (see Section \ref{sec:results}).}
\label{fig:moonrise-scatter}
\end{figure*}

In this section, we present the results obtained by testing the ANN discussed in Section \ref{sec:methods} on our test set (see Section \ref{sec:dataset}). As was previously discussed, \texttt{M-TOPnet} takes as input pairs of normalised continuum-subtracted spectra and normalised ten-point continuum vectors, and it produces the following outputs: 
\begin{itemize}
    \item a 12.217-sized 1D vector indicating the predicted locations of spectral lines reported in Table \ref{tab:spectral_lines};  \newline
    
    \item a 800-sized 1D vector representing the PDF of the redshift prediction, with minimum redshift bins of $dz = 0.003$; \newline
    
    \item a scalar value for the $M_{\rm star}$ prediction; \newline
    
    \item a scalar value for the SFR prediction.
\end{itemize}

We note that we used scalar predictions for $M_{\rm star}$ and SFR because calculating PDFs for these quantities would significantly slow down the training and inference processes due to the increased model complexity and computational demand.
Given the importance of capturing potentially multi-peaked solutions, we prioritised generating a PDF for the redshift prediction. This approach serves as a proof of concept for our method: due to its modular design, our pipeline can easily be extended to produce PDFs for $M_{\rm star}$ and SFR in future work.
\\\\
The upper panels of Figure \ref{fig:residuals1} display scatter plots comparing labelled values and predicted values for all 17~729 simulated MOONS galaxy spectra in our test set, considering exposure times of $2~h$, $4~h$, and $8~h$. 
While the comparison for $M_{\rm star}$ and SFR is straightforward as they are scalar values, the redshift comparison is less direct due to the adopted classification scheme where both labels and predictions are probability vectors. For this comparison, we use the peaks of both labelled and predicted PDFs.

Our model demonstrates excellent agreement with the ground truth values, as is evidenced by the red points (representing instances in the test set) aligning closely with the one-to-one relation (black line, representing a perfect model). Notably, redshift predictions are highly accurate, with most points coinciding with the one-to-one relation and only a small fraction of outliers, confirming the effectiveness of training through a classification scheme and handling discretised distributions of continuous values.

The lower panels of Figure \ref{fig:residuals1} present histograms (in violet) showing the distributions of residuals for the relations in the upper panels. These residual distributions, defined as ‘predicted value - labelled value’, are fitted with a Gaussian model (blue curve), and the best-fit values for $\mu$ and $\sigma$ are also shown. For the redshift, we observe $\mu = 0$, indicating no appreciable offset between predicted and labelled redshift. 
$M_{\rm star}$ shows a $\mu = 0.02$ dex, and SFR a $\mu = -0.005$ dex, both demonstrating minimal offset. 
Regarding $\sigma$ values, we find 0.15 dex for $M_{\rm star}$, 0.22 dex for SFR, and a remarkable 0.001 for redshift. 
These results confirm our earlier observations about the upper panels of Figure \ref{fig:residuals1}. 
It is important to note that for redshift, the Gaussian fit primarily represents the bulk of data points near the one-to-one relation and does not account for outliers, which will be discussed in detail in subsequent paragraphs. 

One evident cause of outliers in the redshift determination is the spectrum quality, determined by the exposure time, $t_{\rm exp}$. This is demonstrated in the left panel of Figure \ref{fig:residuals2}, where we restrict the model predictions to mock spectra with a simulated exposure time of 8 hours. This analysis clearly shows that most outliers, especially the most catastrophic ones, disappear ($<2~\%$).

The right panel of Figure \ref{fig:residuals2} presents the same analysis on the test set, but using a version of \texttt{M-TOPnet} where multi-task learning is artificially disabled by assigning a weight of 0 to all losses except the redshift determination one. 
This plot reveals a significant number of outliers, about 5\% of which are catastrophic (with differences between labels and predictions up to $\Delta z > 0.5$), confirming the benefits of multi-task learning in achieving better representations, as is discussed in Section \ref{sec:methods}.
Additionally, when we disable only the $M_{\rm star}$ and SFR tasks, while keeping the line location task active, we still observe catastrophic outliers, with a fraction of about 4\%, indicating that these quantities provide important constraints on the redshift solution space. 
This result highlights the intrinsic physical relationships among these quantities: the continuum, emission lines, $M_{\rm star}$, and SFR are interrelated, and leveraging these connections helps the model make more accurate redshift predictions.
\\\\
To validate our model in a scenario more closely aligned with observational strategies, we conducted a test by applying physically motivated selection criteria to galaxies in our simulated test set. Specifically, we emulated the selection criteria anticipated by the planned GTO MOONS extragalactic survey MOONRISE (\citealp{Maiolino2020}).
For this analysis, we first selected spectra of galaxies at $z < 2.6$ with $m_H < 25$. 
For the latter criterion, we adopted a conservative approach by including fainter objects; however, this is still conservative, as the survey is expected to reach down to $m_H = 24$ (\citealp{Maiolino2020}).
Among these selected spectra, we further refined our selection based on the galaxy type and exposure time: spectra with simulated $t_{\rm exp} = 2~h$ for star-forming galaxies and $t_{\rm exp} = 8~h$ for passive galaxies (see Figure \ref{fig:dataset_1}b; see also a thorough discussion on the justification of such selection criteria in \citealp{Maiolino2020}). 
Our MOONRISE-like sample comprise 5~344 objects. Figure \ref{fig:moonrise-scatter} presents scatter plots comparing labelled values and predicted values for redshift, $M_{\rm star}$, and SFR for this selected subset of the test sample. These plots show similar trends to those observed in the general case, as the applied selection criteria do not significantly alter the bulk properties of the simulated galaxy population.
This similarity is also evidenced by the distributions of residuals, which closely resemble those of the full dataset. The $\mu$ and $\sigma$ of these distributions, obtained by fitting Gaussian models, are displayed in the lower right corners of each panel in Figure \ref{fig:moonrise-scatter}.
We provide additional quantitative analysis of \texttt{M-TOPnet} performance on the MOONRISE-like sample in the next sections, including specific metrics and comparisons to the full dataset results.

\begin{figure}[ht!]
\centering
\includegraphics[width=1\columnwidth]{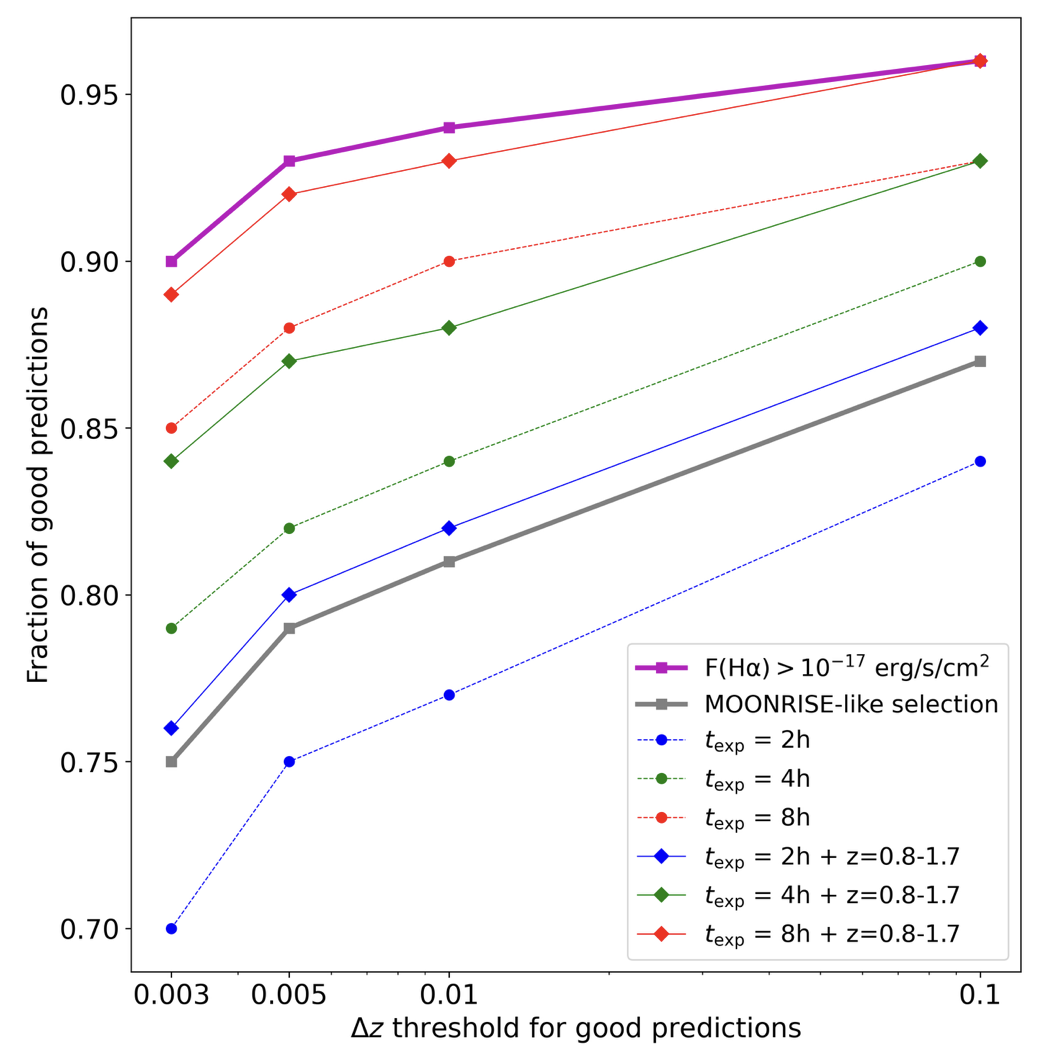}
\caption{Fraction of successful redshift predictions as a function of varying $\Delta z$ thresholds for different galaxy subsets. The subsets include a general division by exposure time ($t_{\rm exp} = 2~h, 4~h, 8~h$, represented by dashed lines) and a more specific division by both exposure time and redshift range $z = 0.8-1.7$ (represented by solid lines in blue, green, and red). The MOONRISE-like selection is shown by the solid grey line, and bright galaxies with $H\alpha$ flux greater than $10^{-17} \rm ~erg/s/cm^2$ are shown by the solid violet line. The $\Delta z$ thresholds tested are 0.003, 0.005, 0.01, and 0.1.}
\label{fig:successRate}
\end{figure}

\subsection*{Spectroscopic redshift determination}

From this point forward, we concentrate our analysis on the redshift predictions. 
This focus is warranted because redshift prediction is a particularly challenging task, as it relies heavily on complex information from the full spectrum. Additionally, achieving high accuracy in redshift predictions is crucial for most of the relevant measurements (see a discussion in Section \ref{sec:intro}). 
Importantly, \texttt{M-TOPnet} employs novel and sophisticated methods for redshift determination, which could potentially become standard in astronomical ML pipelines.

While previous analyses focussed on the fraction of test galaxy spectra for which redshift was ‘exactly’ determined (i.e. the peak of the predicted redshift PDF is in the same $dz=0.003$ sized bin as the peak of the target redshift), another useful diagnostic is to determine the fraction of ‘successful predictions’ given varying thresholds of $\Delta z$. Figure \ref{fig:successRate} visualises this test using four increasing $\Delta z$ thresholds: 0.003, 0.005, 0.01, and 0.1 (the last being consistent with typical photometric redshift uncertainties).

Here, it is crucial to note that the total dataset includes varying levels of noise (exposure times from $2~h$ to $8~h$) and a wide distribution of physical properties (see Section \ref{sec:dataset}), including quenched or green-valley galaxies with faint or absent emission lines, making accurate prediction challenging for some spectra: in Section \ref{sec:discussion}, we shall expand on some crucial differences in terms of physical properties between successful and unsuccessful predictions.

Thus, we analysed the sample by splitting it by exposure time (dashed blue, green, and red lines for $2~h$, $4~h$, and $8~h$ observations, respectively), confirming that indeed mock galaxy spectra with a longer simulated exposure time reach higher fraction of successful predictions, ranging from about 0.85 at $\Delta z < 0.003$ up to 0.92 at $\Delta z < 0.1$.
In grey (solid line), we present the trend for the subset of galaxies defined by selection criteria similar to those of the MOONRISE GTO survey (see the previous section). This subset exhibits approximately 76\% successful predictions at a $\Delta z$ threshold of 0.003, increasing to about 85\% at a $\Delta z$ threshold of 0.1. 
We also selected test galaxy spectra in the redshift range $z = 0.8-1.7$ (corresponding to about 60\% of the test set)
without any additional selection criteria: this is an interesting redshift range as it corresponds to the one at which most of the observations from MOONRISE have been planned (\citealp{Maiolino2020}), 
and offers an ideal target range for all future surveys. 
In fact, at these redshifts the H$\alpha$ line falls above 1.2$\mu$m, maximising the impact of the unique MOONS H-band channel. In addition, this range allows for simultaneous coverage of bright optical emission lines like the [OII] doublet at around 3728 $\AA$, and the $H\beta$ and the [OIII] lines at 4960 $\AA$ and 5008 $\AA$, respectively. 
Focussing on this redshift range results in an approximately 5\% improvement in successful predictions (see solid blue, green, and red lines).
Lastly, we also examined ‘bright’ sources with $H\alpha$ flux greater than $10^{-17} \rm ~erg/s/cm^2$ (solid violet line), 
a representative threshold  close to the median value of the $H\alpha$ flux distribution (see Figure \ref{fig:dataset_1}). 
We find that these sources, irrespective of $t_{\rm exp}$, show very high levels of successful predictions, ranging from 90\% at $\Delta z<0.003$ to 95\% at $\Delta z<0.1$.
\\
\begin{figure}[ht!]
\centering
\includegraphics[width=1\columnwidth]{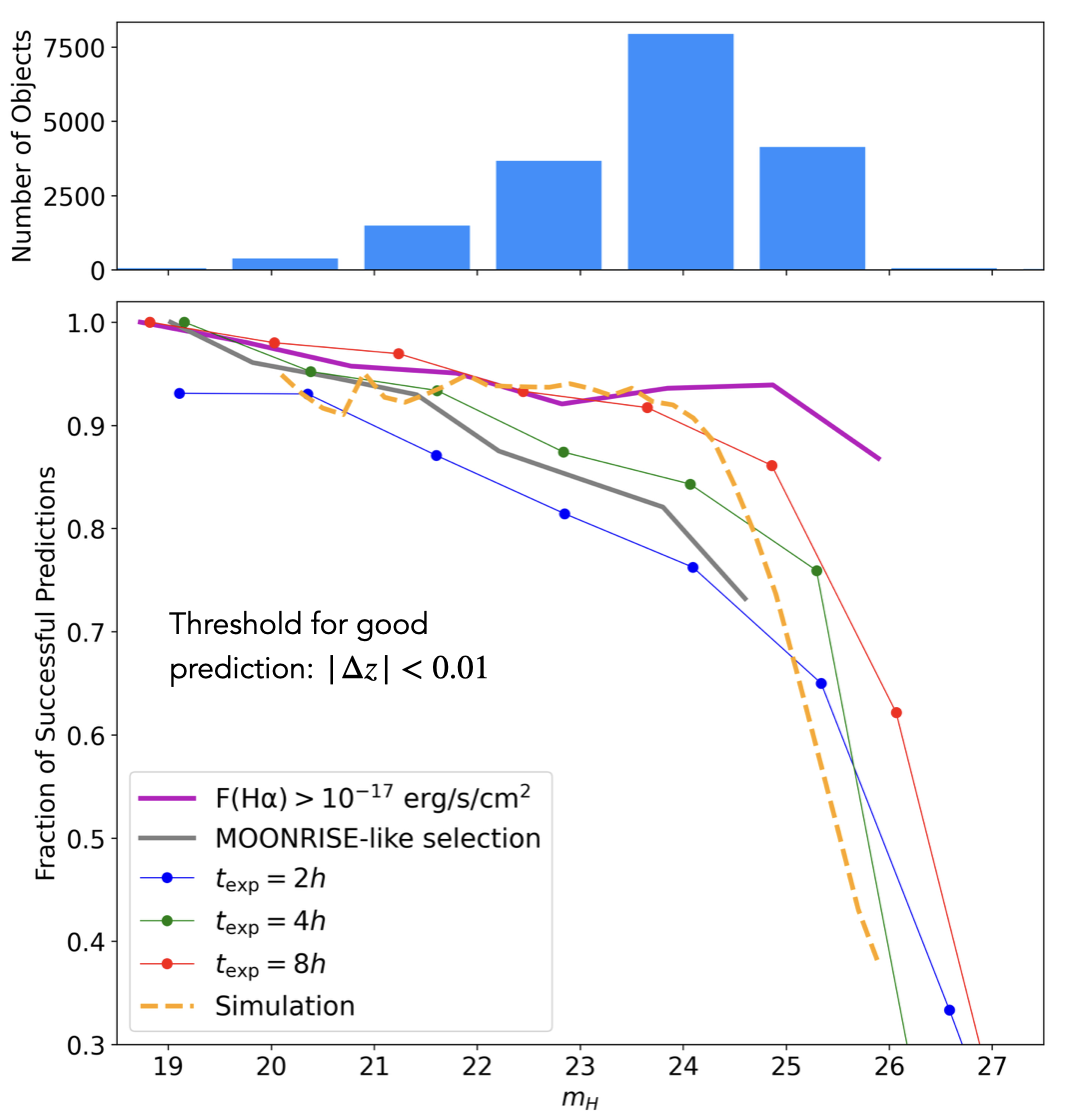}
\caption{Relationship between the fraction of successful redshift predictions (with $|\Delta z| < 0.01$) and the H-band magnitude $m_H$. 
Top panel: Histogram of the number of objects as a function of $m_H$. 
Bottom panel: Fraction of successful predictions as a function of $m_H$ for different subsets: galaxies with $H\alpha$ flux greater than $10^{-17} \rm ~erg/s/cm^2$ (violet line), MOONRISE-like selection (grey line), and subsets with varying exposure times of $2~h$ (blue), $4~h$ (green), and $8~h$ (red), along with a simulated reference trend (dashed orange line). See Section \ref{sec:results} for a detailed discussion on the behaviour of the observed trends.} 
\label{fig:successRate_vs_magH}
\end{figure}

\begin{figure*}[ht!]
\centering
\includegraphics[width=1\textwidth]{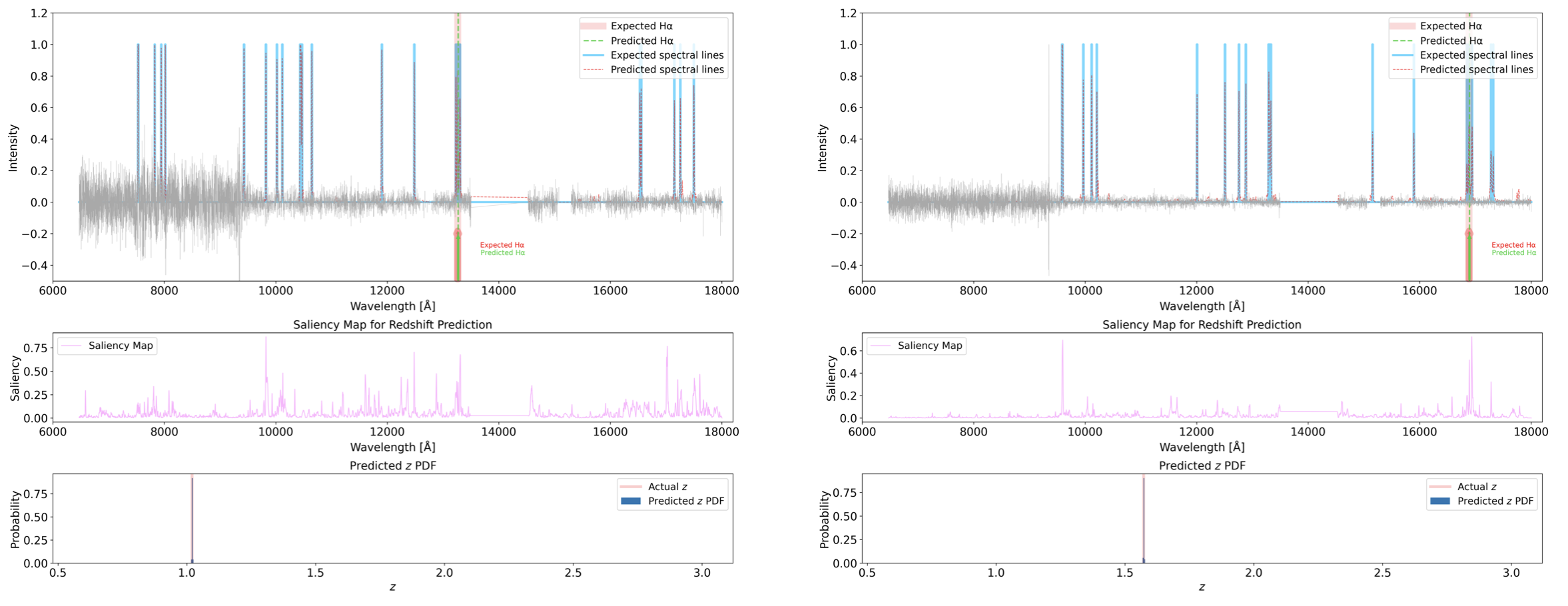}
\caption{Examples of pipeline outputs for two random galaxy spectra with successful redshift predictions ($|\Delta z| < 0.003$). 
Top panels: Input spectra (grey) with predicted line locations (dashed red lines) and expected spectral lines (light blue lines). The expected and predicted locations of the H$\alpha$ line are marked with a red and green arrow, respectively. 
Middle panels: Saliency maps for redshift prediction, indicating the most influential wavelengths for the model predictions. 
Bottom panels: Predicted (blue) and actual (red) redshift PDFs, highlighting the accuracy of the redshift predictions.}
\label{fig:good_pred}
\end{figure*}

\begin{figure*}[ht!]
\centering
\includegraphics[width=1\textwidth]{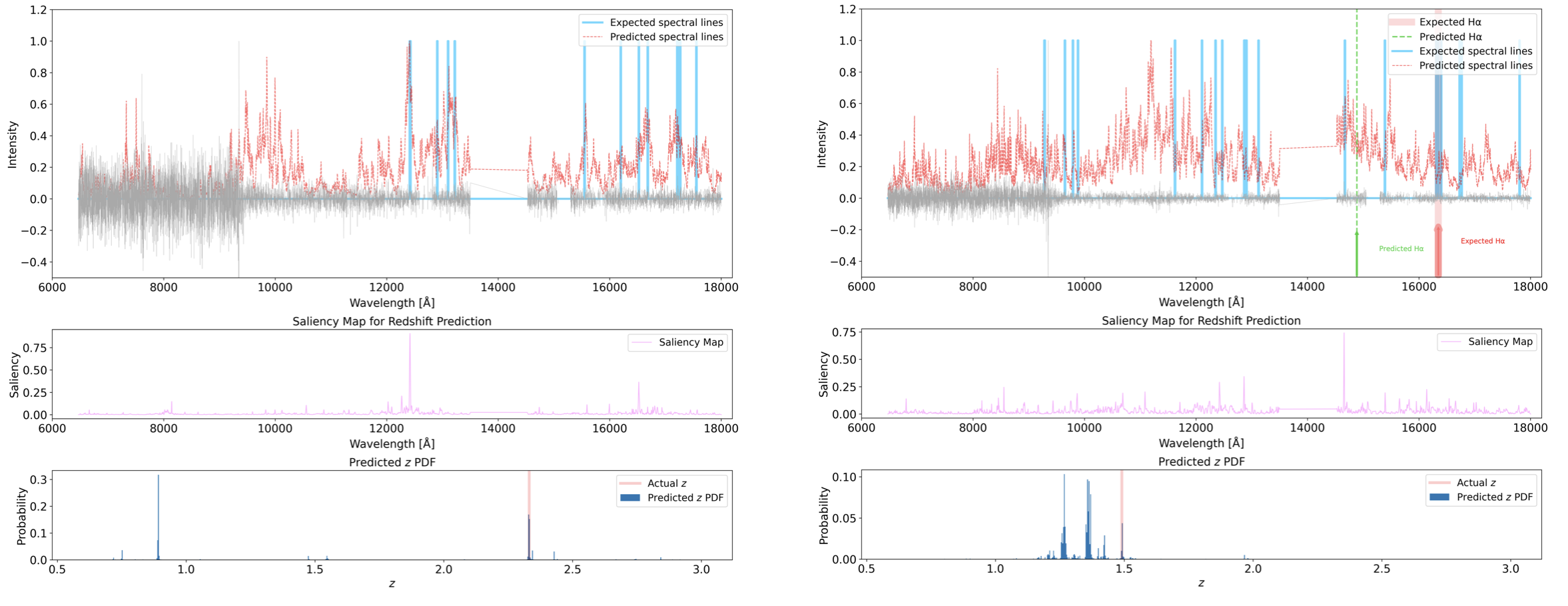}
\caption{Same as Figure \ref{fig:good_pred}, but showing the pipeline outputs for two random galaxy spectra with unsuccessful redshift predictions (see Section \ref{sec:results}).}
\label{fig:bad_pred}
\end{figure*}

\begin{figure}[ht!]
\centering
\includegraphics[width=1\columnwidth]{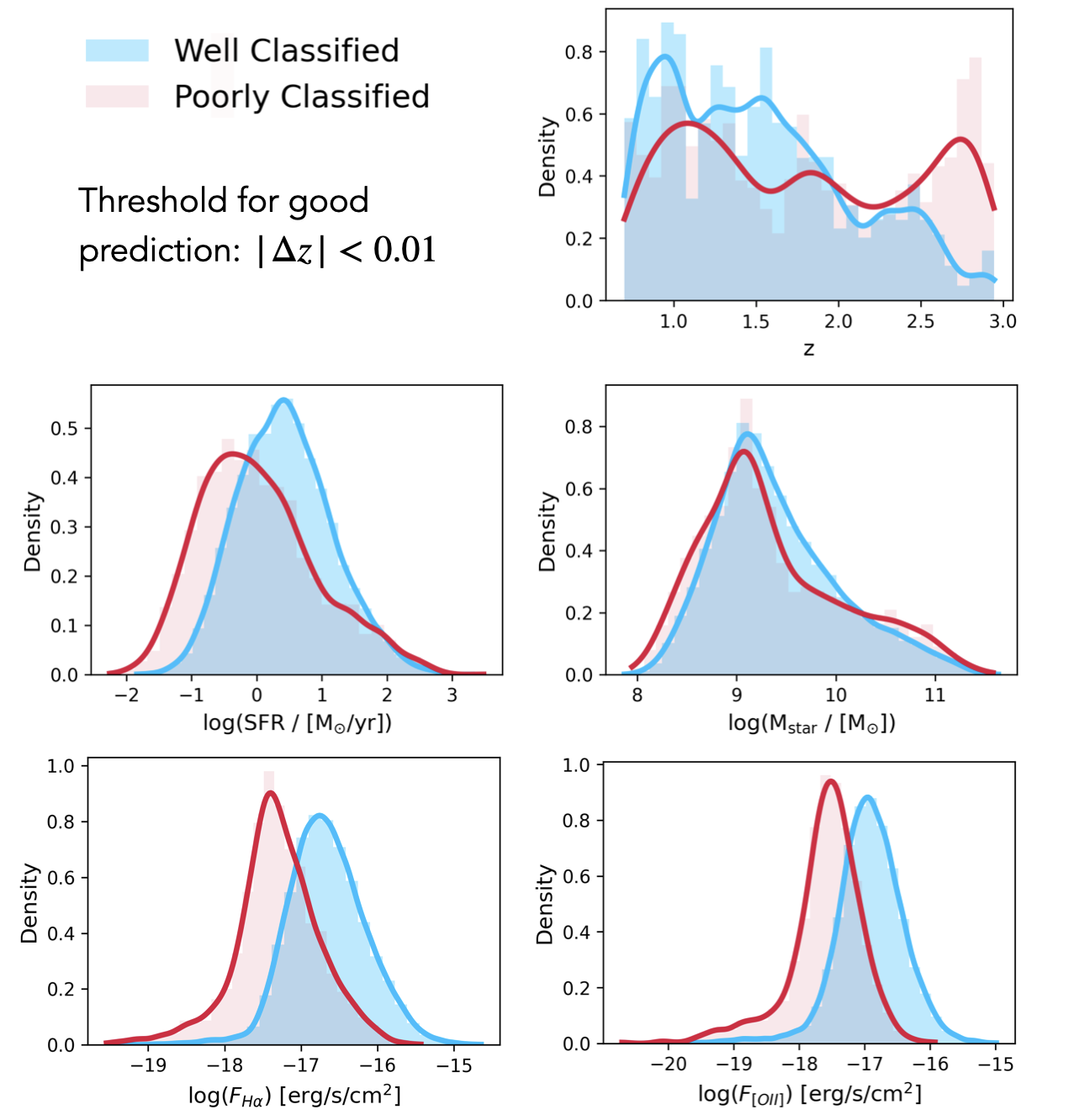}
\caption{Normalised distributions of well-classified (blue) and poorly classified (red) galaxy spectra based on various parameters (see Section \ref{sec:discussion} for a discussion). Top left to bottom right: Distributions as a function of redshift ($z$), SFR, stellar mass ($M_{\rm star}$), and fluxes of $H\alpha$ and [OIII] emission lines. The threshold for a good prediction is set at $|\Delta z| < 0.01$.}
\label{fig:good_vs_bad}
\end{figure}

Another useful check is to determine how the fraction of successful predictions, at a fixed $\Delta z$ threshold, depends on critical  observables that could be used to select target observations for MOONS. 
In Figure \ref{fig:successRate_vs_magH}, we visualise this check using a $\Delta z$ threshold of 0.01 and showing the fraction of successful predictions as a function of $m_H$. We adopted a $|\Delta z| < 0.01$ threshold because it represents an acceptable uncertainty for most of the $z$-based measurements, and we chose $m_H$ due to its wide availability for most likely MOONS targets.
As in Figure \ref{fig:successRate}, we display the MOONRISE-like sample (solid grey lines) and different sub-samples based on exposure times (solid blue, green, and red lines for $2~h$, $4~h$, and $8~h$ exposures, respectively) and $H\alpha$ flux (solid violet line). For very bright galaxies, $m_H$ < 22, the fraction of successful predictions is almost always above 90\%, except for simulated $2~h$ exposures, which reach about 85\% at $m_H$ = 21.5. Between $m_H$ 23 and 25, where most test galaxies lie (see histogram in the top panel), the gap between sub-samples increases: bright $H\alpha$ galaxies and $8~h$ exposures consistently achieve above 90\%, 
while the MOONRISE-like sample is positioned approximately between the 4-hour exposures, showing success rates of 80-90\%, and the 2-hour exposure, ranging from 75-85\%.
At $m_H$ > 25, success fractions drop below 50\%, although bright $H\alpha$ galaxies and $8~h$ exposures still achieve about 90\% and 65\% success rates, respectively, down to $m_H$ $\sim$ 26. 

For reference, the dashed orange line represents the results from a direct computation of the expected detection rate of star-forming galaxies based on observed properties of the COSMOS galaxies. In brief, 
COSMOS galaxy properties are derived by SED fitting from the
COSMOS2020 catalogue \citep{Weaver2021}. The flux of the H$\alpha$ line was derived from the measured SFR and dust extinction, assuming standard calibrations by \cite{Kennicutt2012}.  
The metallicity was estimated from the galaxy stellar mass and SFR, assuming that all the galaxies in the sample  follow the fundamental metallicity relation (\citealp{Mannucci2010}), which is observed to be in place at least up to z$\sim$3 (\citealp{Cresci2019}). At each metallicity, line ratios were inferred using the calibrations presented in \cite{Curti2020}. 
Galaxy sizes, important to estimate the fiber losses, were derived as a function of stellar mass and redshift from \cite{VanderWel2014}. Finally, line widths were derived from the observed Tully-Fisher relation at z$\sim$1 from \cite{DiTeodoro2014}. A redshift was considered to be successfully determined if at least two emission lines were well detected, one with $S/N>7$ and the other with $S/N>5$.
Further details on these computations will be provided in an upcoming work. 
It can be seen that the results from the ML model are in good agreement with the expectations up to $m_H$ $\sim$ 25, and are above these values for fainter magnitudes. 
\\\\
Finally, in Figures \ref{fig:good_pred} and \ref{fig:bad_pred}, we showcase four examples of the output retrieved from \texttt{M-TOPnet} for four galaxy spectra in the test set. 
Figure \ref{fig:good_pred} details two random examples of spectra for which we achieved a successful redshift prediction with the highest possible accuracy ($|\Delta z| < 0.003$). 
For both spectra, shown in grey in the upper panels, the redshift PDF is extremely narrow, and peaked at the location of the expected label redshift. 
Additionally, the line location task performed very well, accurately predicting the location of all the spectral lines defined in the spectral line masked vector (vertical light blue lines), as is indicated by the vertical dashed red lines. As was discussed in previous paragraphs, similar results are obtained for most spectra in the test set.

Figure \ref{fig:good_pred} also includes 1D saliency map vectors for the redshift prediction task (see central panels), obtained by computing the gradients of the model predictions with respect to the input spectrum tensor.
A saliency map highlights the parts of the input that are most influential in determining the output. For our 1D spectra, the saliency map shows which wavelengths are most important for the redshift prediction. We note that the saliency map vectors peak not only at spectral positions where lines are present but also at positions where there are no lines. This suggests that the model also considers the absence of lines at certain positions as informative for determining the redshift, as the presence of a line in these locations would change the solution.

Figure  \ref{fig:bad_pred} shows two random examples of unsuccessful redshift predictions, characterised by wider redshift PDFs with multiple peaks and failure in the line location task. Interestingly, in the first case (left panel), the secondary solution predicted by our model corresponds to the correct one, whereas this is not the case in the second example (right panel). By examining and visualising various outputs from the test set, we realise that the small fraction of spectra with large errors tend to have wider redshift PDFs, which become almost flat with numerous tiny peaks in instances of catastrophic errors ($|\Delta z| > 0.5$).
This behaviour can be understood from a Bayesian perspective: when the spectral data provides insufficient evidence to strongly constrain the likelihood, the posterior distribution becomes broader and more uncertain, reflecting the lack of strong evidence to favour a single redshift solution. This behaviour highlights the ability of the model to express uncertainty in its predictions under ambiguous conditions.
\\\\
As was mentioned in Section \ref{sec:methods}, we also experimented with MC dropout for redshift determination, by averaging results from 50 forward passes with dropout layers active during inference. This technique addresses epistemic uncertainty by approximating a Bayesian neural network (see Section \ref{sec:intro} and \ref{sec:methods} and references therein). 
We find that MC dropout does not provide a tangible improvement in the global results discussed above and shown in Figures \ref{fig:residuals1}, \ref{fig:successRate}, and \ref{fig:successRate_vs_magH} (values remain similar within a relative factor of about 10\%). 
This may be because the technique of discretising the redshift range and turning its prediction into a classification task accounts for most of the uncertainty, which thus appears to be primarily aleatoric rather than epistemic in this problem. 
However, by visually inspecting the difference between the averaged prediction obtained through MC dropout and the prediction from a single inference, we observe an appreciable net effect in a few cases. 
In these cases, the label redshift corresponds to the secondary solution of the predicted PDF in the single-inference case, and the situation improves with more forward passes, leading to a correct prediction.
Thus, while the global effect of MC dropout is minimal, it has the potential to improve results on an individual case basis for certain types of spectra. 
This indicates potential, and we plan to conduct a more detailed examination of this technique in future work to identify the spectra classes for which it can have a more significant impact and determine the optimal balance between the number of forward passes and total computation time.

\section{Discussion}
\label{sec:discussion}

The results presented in Section \ref{sec:results} clearly demonstrate the potential of ML in inferring information from spectral data of galaxies. In particular, our \texttt{M-TOPnet} pipeline has proven highly effective in accurately determining galaxy physical properties. These findings serve as a valuable benchmark for planning future observational strategies with MOONS.

\subsection{Physical properties of galaxies with incorrectly predicted redshifts}

We have conducted several analyses to test the precision of \texttt{M-TOPnet} for redshift determination. These tests reveal that a small fraction of test galaxy spectra (varying with exposure time) are not accurately predicted. This observation raises an intriguing question: whether we can identify and understand the causes of these prediction failures, and furthermore, whether these failures are driven by inherent properties of the data rather than model misspecification.

These questions connect to the broader topic of explainability in deep learning, which has become a central focus in ML research. 
Explainable ML aims to make the decision-making processes of complex models more transparent and interpretable. In the context of our study, understanding where and why our model fails can provide valuable insights into its limitations and potential areas for improvement.

To address these questions, we analysed the differences between the distributions of simulated MOONS galaxy spectra in the test set that are successfully predicted versus those that are not. We adopted the same threshold on the residuals of $|\Delta z| < 0.01$, as in Section \ref{sec:results}, to define successful predictions.

Figure \ref{fig:good_vs_bad} illustrates the distributions of both sub-samples. To facilitate a fair comparison, given the significantly smaller number of poorly determined redshifts, we normalised each distribution to the same area. The histograms show the distributions of successfully predicted objects in blue and unsuccessfully predicted objects in red, as a function of redshift, SFR, $M_{\rm star}$, and fluxes of $H\alpha$ and [OIII] emission lines.
We observe clear offsets between the red and blue distributions in the case of SFR and emission line fluxes. The histograms of unsuccessful predictions peak at systematically lower SFR and emission line fluxes (similar behaviours are observed for other emission lines, provided in the simulated dataset). This finding is not surprising, as galaxies with lower SFR and lower emission line fluxes have fainter (or absent) spectral lines, presenting a challenge for accurate redshift determination. 
The distributions of $M_{\rm star}$ show less obvious deviations; however, we note a relative excess of unsuccessful predictions at very low $M_{\rm star}$ ($<10^9 \rm M_{\odot}$), where galaxies tend to have very low SFR according to the galaxy main-sequence, and at very high $M_{\rm star}$ ($>10^{10.5} \rm M_{\odot}$), where a non-negligible fraction of the galaxy population is quenched, as is seen in the $M_{\rm star}$-SFR diagram (see Section \ref{sec:dataset}).

Another noteworthy finding emerges from the analysis of the redshift distributions. We observe a significant relative excess of unsuccessful predictions at redshifts between $z=2.7-3$, also visible in the scatter plots of Figure \ref{fig:residuals2} (see the grey squared box delineating the redshift range under examination). 
This behaviour can be explained by the simultaneous absence of bright Hydrogen and Oxygen emission lines, which in this redshift range fall outside the spectrum or in the gap between spectral bands.

\subsection{Analysis and screening of the output redshift probability distribution functions}

Having identified differences in physical properties between galaxy spectra with successfully and unsuccessfully determined redshifts, we now turn our attention to potential differences in the pipeline outputs themselves.
\texttt{M-TOPnet} does not output a single scalar value for redshift but rather a probability vector. This raises the question of whether  
the shape of the PDF can be used to derive the quality of the prediction and, in particular, to identify the wrong redshift estimates.
Intuitively, one might expect that galaxy spectra with unsuccessfully predicted redshifts (for instance, due to the reasons discussed above and visualised in Figure \ref{fig:good_vs_bad}) would have broader or more dispersed PDFs with multiple peaks.
This can be justified in the context of the Bayes' theorem, as insufficient evidence in the spectral data leads to a poorly constrained likelihood, resulting in a dispersed posterior.

\begin{figure}[ht!]
\centering
\includegraphics[width=1\columnwidth]{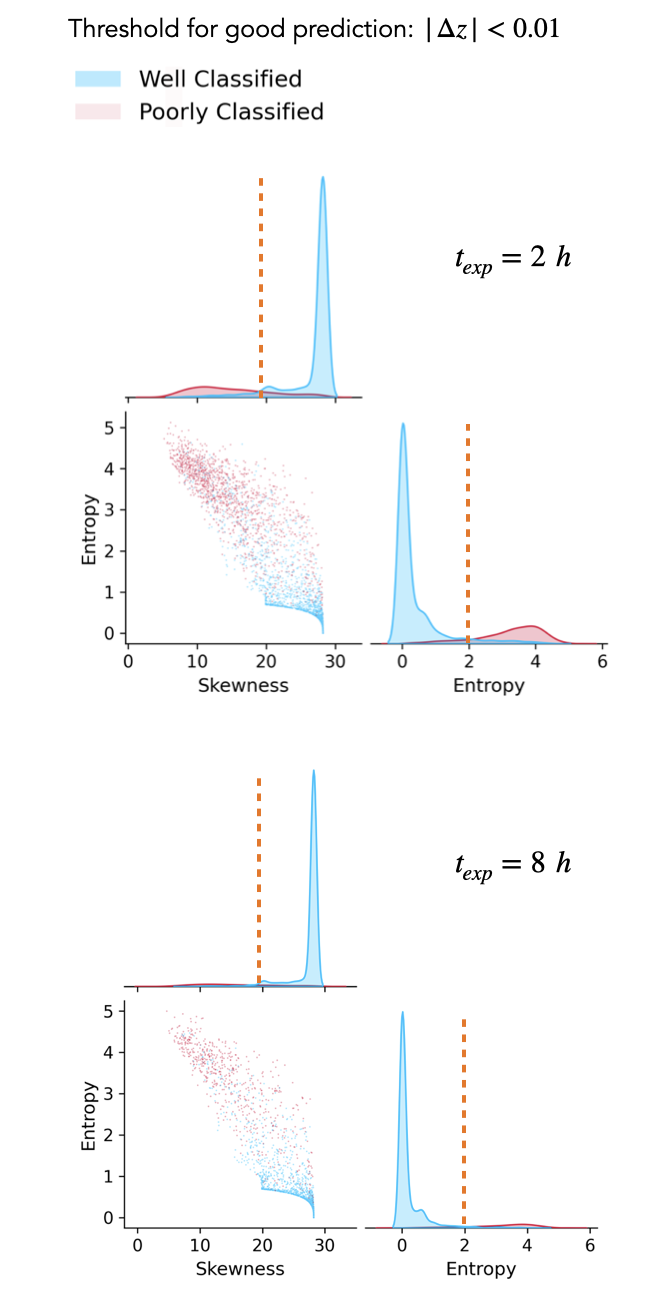}
\caption{Corner plots showing the distributions and mutual relations of skewness and entropy of redshift PDFs for well-classified (blue) and poorly classified (red) galaxy spectra. The threshold for a good prediction is set at $|\Delta z| < 0.01$.
Top panel: Results for spectra with an exposure time of $t_{\rm exp} = 2~h$. 
Bottom panel: Results for spectra with $t_{\rm exp} = 8~h$. 
Well-classified spectra and poorly classified spectra are clearly distinguished by their skewness and entropy values (see a discussion in Section \ref{sec:discussion}). 
The vertical orange lines indicate the separation threshold for skewness $= 20$ and entropy $= 2$.}
\label{fig:checkPDF}
\end{figure}

The challenge lies in quantifying this intuition objectively, potentially enabling further a posteriori screening of the output. 
To this end, we employed \texttt{scipy.stats} methods to compute the entropy 
and the skewness of the redshift PDFs obtained for the galaxy spectra in our sample. 
Entropy, in this context, quantifies the uncertainty or spread in the redshift probability distribution, indicating how concentrated or dispersed the predicted redshift values are.
Skewness, on the other hand, measures the asymmetry of the redshift PDFs. In our context, we expect poorly classified spectra to produce broader, flatter PDFs with lower peak amplitudes, leading to skewness values closer to zero, while well-classified spectra should result in narrow, asymmetric PDFs with higher skewness values.
Figure \ref{fig:checkPDF} presents the distributions and mutual relations of these metrics in corner plots, maintaining the colour coding from Figure  \ref{fig:good_vs_bad} (red for poorly predicted redshifts, blue for successfully predicted spectra). For this analysis, we divide the test set into its two extreme exposure times, showing objects with $t_{\rm exp}=2~h$ in the top panel and objects with $t_{\rm exp}=8~h$ in the bottom panel.
Interestingly, we find that the distributions are double-peaked, and the blue and red distributions are clearly separated in both proxies: poorly predicted objects show smaller skewness and larger entropy, with only a tiny fraction of false negatives (i.e. spectra for which the redshift is successfully determined but whose PDFs appear similar to those of poorly determined spectra). 

The clear separation exhibited in the distributions, which we identify to be around skewness $= 20$ and entropy $= 2$, allows for the following experiment. By using a threshold of entropy $= 2$ (see orange vertical lines in Figure \ref{fig:checkPDF}), we can separate a posteriori – that is, after the process of inference and after obtaining the predictions – the bulk of good and poor predicted spectra. Applying this threshold, we find that the accuracy in redshift determination (using the usual $|\Delta z| < 0.01$ threshold) increases from 0.77 to 0.97 in the $t_{\rm exp}=2 h$ case, and from 0.9 to a remarkable 0.99 in the $t_{\rm exp}=8 h$ case. This improvement corresponds to filtering out about 27\% of the spectra in the first case and about 12\% in the second case.

As is visible in Figure \ref{fig:checkPDF}, this process also rejects a small fraction of false negatives. However, if the specific kind of analysis is not adversely affected by filtering out a fraction of spectra of the order of the percentages reported, one can decide to apply this further screening to strongly enhance the model's performance.
We note that applying a threshold at skewness $= 20$ yields equivalent results, indicating that these metrics independently serve as effective proxies for distinguishing between well- and poorly predicted spectra. While one could consider adopting a threshold based on combinations of these metrics, this goes beyond the scope of this paper, and such a combined approach would likely be less interpretable.
\\
We emphasise that this approach is only possible because \texttt{M-TOPnet} outputs probability vectors rather than single scalar values.

\subsection{Exploration of the embedding space structure}

Another central subject in ML is the interpretability of models, which involves understanding how a model achieves its tasks and what kind of information it uses and encodes. 
Interpretability is thus crucial for validating the decisions of a model and ensuring its reliability. 
One way to address this problem is to study the structure of the embedding spaces. 
The embedding space is a lower-dimensional representation of the input data learned by the model, which captures essential features for the task at hand. 
Several works in astronomy have explored this approach by visualising a dimensionality-reduced structure of key layers in their models (\citealp{Portillo2020, Pat2022, Liang2023, Stoppa2023}), for instance using techniques like UMAP (uniform manifold approximation and projection; \citealp{McInnes2018}) or t-SNE (t-distributed stochastic neighbour embedding; \citealp{Laurens2008}) to analyse latent spaces and gain insights into how these models organise and interpret data (see e.g. \citealp{Sarmiento2021, Melchior2023}).

\begin{figure*}[ht!]
\centering
\includegraphics[width=1.1\textwidth]{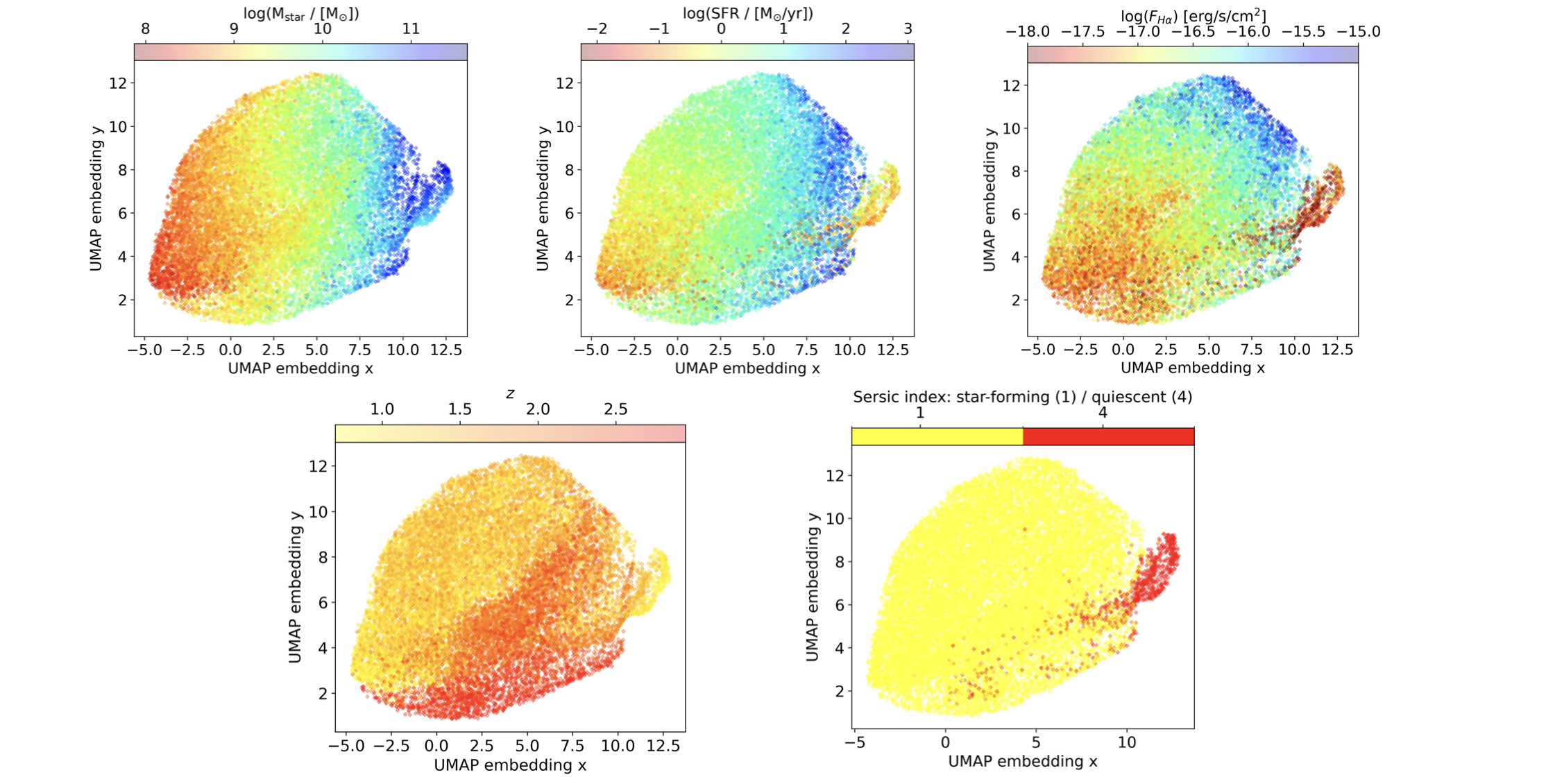}
\caption{UMAP-reduced 2D visualisations of the concatenated dense layers following the ‘shared info’ layer in \texttt{M-TOPnet} (see Figure \ref{fig:nn-sketch} and a discussion in Section \ref{sec:discussion}). 
Each panel shows the embedding space colour-coded by different parameters: from left to right, $M_{\rm star}$, SFR, $H\alpha$ flux, and Sérsic index, which differentiates between star-forming galaxies (Sérsic index = 1) and quiescent galaxies (Sérsic index = 4).}
\label{fig:umap}
\end{figure*}

Following a similar approach, we explored the structure of the concatenation of the three dense layers that follow the ‘shared info’ layer, from which the tasks for redshift, $M_{\rm star}$, and SFR determination depart (see Figure \ref{fig:nn-sketch}). The concatenation of these layers has a dimension of 64x3, but we reduced this to two dimensions for simpler visualisation using the UMAP algorithm. UMAP works by preserving the global structure of the data while maintaining the local relationships between points, making it an effective tool for visualising complex high-dimensional data (\citealp{McInnes2018}).

Figure \ref{fig:umap} shows the UMAP-reduced 2D visualisations of these embedding layers, colour-coded by $M_{\rm star}$ (top left), SFR (top centre), $H\alpha$ flux (top right), redshift (bottom left), and Sérsic index (bottom right). 
The Sérsic index, provided by the simulation, is $n = 1$ for star-forming galaxies and $n = 4$ for quiescent galaxies.

We find that the embedding space under examination successfully encodes information on these quantities, as is evidenced by clear trends and separation of colours. 
This occurs even for quantities like line fluxes (we tested other lines besides $H\alpha$, finding similar results) and star-forming or quiescent classes that were not seen during training and for which the model has no prior knowledge. This suggests that the model is building this knowledge by encoding all the necessary information from the spectral data.
This finding is particularly significant as it demonstrates that, while our model may still function as a black box, it is learning to encode physically meaningful and interpretable information about galaxy properties. By correctly organising galaxies in the embedding space according to their physical properties – even those not explicitly used in training – the model shows that it has developed a meaningful internal representation of the relationship between spectral features and galaxy characteristics.
We note that the redshift trend in the embedding space appears to be less smooth compared to other quantities. This is likely a result of the discretised nature of the task, which introduces step-like transitions in the latent space and affects continuity; this is why we used a more perceptually uniform colour map to better highlight the organisation of the data.

These results collectively demonstrate that our method, in addition to being fast and accurate, is also 
interpretable. 
This further emphasises the potential of well-designed deep learning pipelines for galaxy spectral fitting. 
By providing not just accurate predictions but also insights into the underlying data structure and relationships, such methods can become powerful tools in advancing our understanding of galactic properties and evolution.
Also, the latent space, beyond encoding physically meaningful information about galaxy properties, can also be utilised for outlier detection (see e.g. \citealp{Liang2023}).

\subsection{Alignment with real data}

A final crucial point of discussion concerns the likely possibility that the current version of \texttt{M-TOPnet} may not perform as well on real data as it does on the simulated data it has been trained on, despite the simulations being designed to best reproduce real data. 
This phenomenon is well-known in ML as a domain gap between a source domain (in our case, the simulations) and a target domain (in our case, the real data).
Domain gaps can be triggered by unavoidable offsets between the source and target distributions and are amplified when the model strongly relies on features of the source domain that are different or not present in the target domain. 
Ideally, if labels were available for the target domain, one could re-train or fine-tune the model using them to build more general representations. However, the problem arises when there are no labels, or very few labels, for the target domain. 
For the MOONS data, using labels derived from real data would require processing through standard astronomical tools and therefore transfer the limitations and biases of those tools to our ML pipeline.

In this situation, domain adaptation methods can help reduce the domain gap without prior knowledge of the target domain's labels. Among these techniques, domain-adversarial neural networks (DANNs; \citealp{Ganin2015}) have been explored in astronomy, such as in the identification of galaxy morphology or galaxy mergers or using training data from cosmological simulations or different instruments (e.g. \citealp{Ciprijanovic2020, Ciprijanovi2021, Huertas-Company2024}), or the classification of ionised nebulae \citep{Belfiore2024}.
DANNs work by forcing the model to learn domain-invariant features that are common between source and target domains. This is typically achieved by incorporating a domain classifier into the network architecture. During training, the model learns to maximise the main task accuracy (e.g. redshift prediction) while simultaneously minimising the domain classifier's ability to distinguish between source and target domain samples. This adversarial training process encourages the model to focus on features that are informative for the main task but invariant across domains.
The application of such domain adaptation techniques could potentially enhance the robustness of \texttt{M-TOPnet} when applied to real MOONS data. It would allow the model to leverage the wealth of information in the simulated data while adapting to the specific characteristics of the observed spectra. This approach could be particularly valuable in the early stages of MOONS operations, enabling rapid and accurate analysis of the first batches of real data.

\section{Conclusions}
\label{sec:conclusions}

The advent of large spectroscopic surveys has revolutionised our understanding of galaxy evolution. 
However, the huge volume of data that will be produced by next-generation instruments like MOONS (\citealp{Cirasuolo2020, Maiolino2020}) presents significant challenges to data analysis and interpretation (see Section \ref{sec:intro} and references therein) as traditional spectral fitting techniques, while accurate, can be time-consuming and computationally expensive (\citealp{Hutchinson2016, Bautista2018, Napolitano2023, Zhong2024}). 

This study explores the potential of ML to efficiently and accurately extract key physical properties from galaxy spectra. 
Our work aims to develop a robust pipeline that can handle the complexities of spectral data while providing interpretable results, in preparation for the wealth of data expected from MOONS observations.

Our approach leverages a multi-task learning framework (\citealp{Caruana1997, Ruder2017, Crawshaw2020}), developing a neural network architecture, dubbed \texttt{M-TOPnet}, which simultaneously predicts redshift, $M_{\rm star}$, SFR, and performs spectral line location. 
This integrated approach proves beneficial, improving overall performance, particularly for redshift determination. 
A key innovation in our method is the treatment of redshift prediction as a classification problem (see similar applications in e.g. \citealp{Carrasco2013, Stivaktakis2018, Stewart2022}). By discretising the redshift range, we achieve high accuracy in redshift determination, while also enabling the output of PDFs rather than single scalar values.

When applied to simulated MOONS data, our pipeline demonstrates a promising performance. 
For a subset of galaxies defined by selection criteria similar to those of the MOONRISE GTO survey (see Section \ref{sec:results}), we achieve approximately 76\% successful redshift predictions with a stringent threshold of $|\Delta z| < 0.003$, improving to about 85\% when relaxing the threshold to $|\Delta z| < 0.1$. 
The results are even more promising for longer exposure times, with $8~h$ exposures reaching success rates of 85\% at $|\Delta z| < 0.003$ and 92\% at $|\Delta z| < 0.1$. 
Notably, for bright sources with $H\alpha$ flux greater than $10^{-17} ~\rm erg/s/cm^2$, our success rates further improve to 90-97\%.

Our analysis has revealed several factors, in addition to simulated exposure times, that influence the prediction accuracy. We observe that galaxies with lower SFRs and fainter emission lines pose greater challenges for accurate redshift determination. We note a relative excess of unsuccessful predictions at redshifts between $z=2.7-3$,  due to the absence of bright emission lines in the observable spectrum at these redshifts.

To further enhance the reliability of our predictions, we developed metrics to assess the quality of output redshift PDFs, including root mean square, skewness, and entropy. Interestingly, we found that the distributions of these metrics are multi-peaked, with a clear separation between successfully and unsuccessfully predicted redshifts. The PDFs of poorly predicted redshifts are typically narrower, with larger entropy values and smaller skewness. This clear distinction, only possible because \texttt{M-TOPnet}  outputs full PDFs rather than single scalar values, allows for effective a posteriori screening. By applying thresholds on these metrics, we are able to significantly increase redshift determination accuracy, from 77\% to 97\% for $2~h$ exposures, and from 90\% to 99\% for $8~h$ exposures. This additional screening step demonstrates the potential for post-processing to further improve the performance of the pipeline, highlighting the advantages of our probabilistic approach.

A key strength of our approach lies in its interpretability. Through exploration of the embedding space structure, we found that our model successfully encoded information on galaxy properties not explicitly used in training, such as emission line fluxes and galaxy types (star-forming vs quiescent). 
This finding is particularly significant as it demonstrates that the model is learning meaningful representations of galaxy characteristics from spectral features, rather than simply functioning as a black box.

Looking ahead, we acknowledge the potential challenges in applying \texttt{M-TOPnet} to real MOONS data due to the inevitable differences between simulated and observed spectra.
To address this, we discuss the potential application of domain adaptation techniques, such as domain-adversarial neural networks (e.g. \citealp{Ganin2015, Ciprijanovic2020, Ciprijanovi2021, Huertas-Company2023}). 
These methods could enhance the robustness of our tool when applied to real data by learning domain-invariant features, thereby bridging the gap between simulations and observations.
\\\\
In conclusion, our findings demonstrate the significant potential of ML techniques in spectroscopic analysis. By combining accuracy, efficiency, and interpretability, \texttt{M-TOPnet} offers a promising approach for handling the large datasets expected from next-generation instruments like MOONS. As we move forwards, this work provides a valuable foundation for advancing our understanding of galaxy evolution across cosmic time, clearing the path for new insights from the wealth of upcoming spectroscopic data.

\begin{acknowledgements}
F.B acknowledges support from the INAF `Fundamental Astrophysics` programs 2022 \& 2023. 
E.D.T. was supported by the European Research Council (ERC) under grant agreement no. 10104075.
A.M., F.M., F.B. and G.C. acknowledge support from grant PRIN-MUR 2020ACSP5K\_002 financed by European Union – Next Generation EU.
VW and NFB acknowledge Science and Technologies Facilities Council (STFC) grants ST/V000861/1 and ST/Y00275X/1.
\end{acknowledgements}

\bibliographystyle{aa} 
\bibliography{biblio}

\end{document}